\documentclass[reprint,preprintnumbers,amsmath,amssymb,nofootinbib,aps,prx]{revtex4-1}

\usepackage{slashed}
\usepackage{amsmath}
\usepackage{amssymb}
\usepackage{amsthm}
\usepackage{hyperref}
\usepackage{indentfirst}
\usepackage{psfrag}
\usepackage{graphicx}
\usepackage[utf8]{inputenc}

\hypersetup{colorlinks=true, linkcolor=blue, urlcolor=blue, citecolor=blue, linktocpage=true}

\begin{document}

\begin{flushright}
{\small INR-TH-2019-008}
\end{flushright}

\title{Review of non-topological solitons in theories with $U(1)$-symmetry }

\author{E. Nugaev$^{1}$}
\email{emin@ms2.inr.ac.ru}
\author{A. Shkerin$^{2}$}
\email{andrey.shkerin@epfl.ch}

\affiliation{$^1$Institute for Nuclear Research of the Russian Academy of Sciences, 60th October Anniversary Prospect, 7a, Moscow, 117312, Russia}
\affiliation{$^2$Institute of Physics, Ecole Polytechnique F\'ed\'erale de Lausanne,  CH-1015 Lausanne, Switzerland}

\begin{abstract}

We provide a review of non-topological solitonic solutions arising in theories with a complex scalar field and global or gauge $U(1)$-symmetry. It covers Q-balls, homogeneous charged scalar condensates, and nonlinear localized holes and bulges in a classically stable condensate. A historical overview is followed by the discussion of properties of solutions, including their stability, from different perspectives. Solitons in models with additional massive degrees of freedom are also revisited, and their relation to one-field Q-balls is showed. We also discuss theories with a gauge field giving rise to gauged Q-balls and theories with dynamical gravity giving rise to boson stars.

\end{abstract}

\maketitle

\section{Introduction}
\label{sec:Intro}

The term ``soliton'' was introduced in physics by Zabusky and Kruskal \cite{Zabusky:1965zz} in 1965. They studied numerically solutions of the Korteveg-de Vries equation modeling pulses in nonlinear dispersive media. They found self-propagating solutions that preserve their shapes and interact elastically with each other. Before, the solutions of the sine-Gordon equation with the same properties were found numerically and analytically by Perring and Skyrme \cite{Perring:1962vs}.\footnote{The exact solution they found was already known by that time.} Soon after, the number of papers about solitons ``entered the regime of exponential growth'' (for reviews of the state of the art in 1970s, see \cite{Scott:1973eg,Makhankov:1978rg}). Today, solitons are one of the cornerstones in many branches of physics, both theoretical and experimental. 

In field theory, soliton means a localized ``lump'' of energy \cite{Coleman:1975qj}. It is a nonlinear particle-like solution of field equations, whose profile is preserved with time and during free propagation, but not necessarily in interactions with other solitons.\footnote{Such solutions are also referred to as solitary waves.} In a broad sense, stability of a soliton is due to the balance between defocusing dispersive and focusing nonlinear effects.

Solitons can be classified as topological and non-topological. The solutions of the sine-Gordon equation mentioned above are of the topological type. These solutions are among the first that were discovered, along with skyrmions \cite{Brown:1995zg}. As for non-topological solitons, their best-known representatives are Q-balls \cite{Coleman:1985ki}. Commonly, the fields in a non-topological soliton approach one and the same of their classical vacuum values at spatial infinity. Hence, the topology of the fields at the spatial boundary is trivial. Note, however, that the latter statement does not necessarily imply the former \cite{Nugaev:2016wyt}. Existence and stability of non-topological solitons is due to symmetries of a theory, or, equivalently, due to a presence of conserved charges. At fixed charges, a soliton provides a local extremum of the Hamiltonian of a theory. If this extremum is a local minimum, the soliton is classically stable.\footnote{Although it may not be absolutely stable, see section \ref{ssec:Balls_stab}. } Otherwise, it is classically unstable.

A soliton corresponds to a stationary point of the Hamiltonian. Generally, this point cannot be reached by a perturbation theory built above the classical vacuum state. In turn, perturbation theory can be built on top of the background solitonic configuration. The obtained spectrum of perturbations is to be quantized in order to incorporate the soliton into quantum theory \cite{Christ:1975wt} (see also chapter 5 of \cite{Rajaraman:1982is}). The treatment of the classical soliton as a leading order semiclassical approximation of a quantum object is valid as long as nonlinear couplings responsible for building the configuration are weak \cite{Coleman:1975qj,Lee:1991ax}. Indeed, consider, for example, a theory with a scalar field $\phi$ and a potential $V(\phi)$, in 1+1 dimensions. Let the potential be such that
\begin{equation}\label{WeakCoupling}
V(\phi)=g^{-2}U(g\phi) \; ,
\end{equation}
where the function $U$ depends on $g$ only via the combination $g\phi$. Then, after the field redefinition $g\phi\rightarrow\phi$, the Lagrangian of the theory gets multiplied by a factor $g^{-2}$, thus justifying the semiclassical approximation in the case when the coupling $g$ is weak. In 3+1 dimensions, the same argument applies as soon as
\begin{equation}\label{WeakCoupling2}
V(\phi)=m^4g^{-2}U(g\phi/m) \; ,
\end{equation}
where $m$ is the relevant mass scale of the theory that can be identified, e.g., with the mass of free boson.

A soliton is a non-perturbative solution of field equations. This means, in particular, that the interaction between the soliton and a test particle is not dominated by any particular particle-particle scattering channel; instead, all channels are important. This is due to the fact that the soliton represents a coherent superposition of particles constituting it. The situation is analogous to non-relativistic Bose-Einstein Condensate (BEC) systems. Excitations above a BEC are well described by the two-body interaction Hamiltonian provided that the condensate is sufficiently dilute. Correspondingly, scattering of a particle off a soliton occurs via two-particle interactions if the energy density of the soliton in the interaction region is sufficiently low.

Consider a theory with a complex scalar field and global or local $U(1)$-symmetry. Then, under certain conditions on a scalar field potential, the theory admits localized stationary solutions of finite charge and energy called Q-balls \cite{Coleman:1985ki} (see chapter 6 of \cite{Shnir:2018yzp} for a recent review). A history of their studies goes back to the pioneering work by Rosen \cite{Rosen}. Q-balls found numerous applications in particle physics and cosmology. On the one hand, they were predicted in supersymmetric extensions of the Standard Model \cite{Kusenko:1997zq,Dvali:1997qv,Kusenko:1997vi}, applied to the problems of baryogenesis \cite{Affleck:1984fy} and phase transitions in the Early Universe \cite{Frieman:1988ut}, considered as a dark matter candidate \cite{Kusenko:1997si}. They have also been proposed as alternatives to black holes in certain ranges of masses. For example, a large self-gravitating and rotating Q-ball in a galaxy center is hardly discriminated from a supermassive Kerr black hole \cite{Vincent:2015xta,Troitsky:2015mda} (see also \cite{Akiyama:2019cqa}). On the other hand, Q-balls are relatively simple to investigate and, hence, they can serve as prototypes of other objects whose dynamics in a realistic setting is complicated. For example, solitonic boson stars reduce to Q-balls in the limit when gravity is decoupled \cite{Lynn:1988rb}. In turn, boson stars can model astrophysical objects, such as neutron stars, in the situations when their internal dynamics is not important; see \cite{Liebling:2012fv} for a review.

Studying Q-balls provides insights into behaviour of other solutions of classical equations of motion. In theories with real scalar fields there is no conserved $U(1)$-charge that would ensure the stability of solitons; yet, such theories can admit long-lived localized periodic solutions called oscillons \cite{Bogolyubsky:1976nx,Gleiser:1993pt}. The latter could form in the Early Universe as a result of fragmentation of the inflaton field after the end of inflation \cite{Amin:2011hj}. Recent studies suggest that oscillons can transfer matter-antimatter asymmetry from the inflaton field to baryons \cite{Lozanov:2014zfa}, that mutual collisions of oscillons can produce gravitational wave signal \cite{Cotner:2018vug,Lozanov:2019ylm}, or they can form primordial black holes \cite{Cotner:2018vug}.

Having constructed Q-balls in a relativistic field theory, one can study systematically their non-relativistic limit. However, the (global) $U(1)$-symmetry, necessary for the existence of solitons, appears naturally in the non-relativistic regime of a theory even if the corresponding relativistic theory does not possess such a symmetry (see, e.g., \cite{Mukaida:2016hwd,Namjoo:2017nia}). Non-relativistic systems in which analogs of Q-balls exist include BEC, see more on this in section \ref{ssec:Motivation}. 

Q-balls are not the only nontrivial classical solutions in complex scalar field theories with a (global) $U(1)$-symmetry. They are necessarily accompanied by a spatially-homogeneous charged scalar condensate.\footnote{By spatially-homogeneous we understand a solution both magnitude and phase of which are independent of position in space. Note that it is possible to construct configurations breaking spatial translations spontaneously but with the linear combination of translation and $U(1)$ generators remain unbroken; see, e.g., \cite{Musso:2018wbv}.} Understanding properties of the condensate is important, for example, when one studies the dynamics of scalar fields after inflation; for a recent review see \cite{Amin:2014eta}. Depending on a scalar field potential, condensate solutions, as well as Q-balls, can be either classically unstable or classically stable \cite{Kasuya:2000wx,Gorbunov:2011zz}. (In)stability of the condensate is relevant for the question of formation of Q-balls and in studies of possible transitions between homogeneous and non-homogeneous configurations. Among other types of nonlinear classical configurations possibly present in a theory one can mention Q-holes and Q-bulges \cite{Nugaev:2016wyt}. These are, accordingly, local nonlinear dips and bumps in the classically stable condensate. Interestingly, the fields of these solitons at spatial infinity do not match with the classical vacuum state. In this review, we discuss in detail the four aforementioned types of solutions.

Q-balls have direct analogs in theories with more than one (complex) scalar field. The noteworthy example is the soliton in the theory of two interacting scalar fields studied by Friedberg, Lee and Sirlin \cite{Friedberg:1976me}. In the case when the relevant mass scales of additional fields are large compared to that of the $U(1)$-field, integrating out heavy degrees of freedom reduces the study of multi-field non-topological solitons to the study of ordinary Q-balls in some effective potential. The case when the hierarchy between the scales is not respected is more challenging. Finally, a special treatment is needed for additional massless fields, gauge fields and gravity. 

Typically, knowing the charge $Q$ and the energy $E$ of a Q-ball is enough to determine it uniquely.\footnote{There can be only a finite number of solutions which differ by the angular velocity but have coincident $E$ and $Q$.} If one also includes the angular momentum $J$ in the list of conserved quantities characterizing stationary solutions, one arrives at spatially localized axially-symmetric spinning Q-balls \cite{Volkov:2002aj} (see also \cite{Kleihaus:2005me}) or elongated vortex-like configurations called Q-tubes (see, e.g., \cite{Sakai:2010ny,Tamaki:2012yk,Nugaev:2014iva}). If the requirement of being stationary is dropped, the amount of classical configurations inflates further, often in exchange for simple analytical treatment. 

This review is aimed at revisiting non-topological solitons in theories with a global or local $U(1)$-symmetry, and at highlighting recent achievements in the field. In section \ref{sec:Gen}, we recapitulate from different perspectives general properties of classically stable Q-balls, classically unstable Q-balls (also referred to as Q-clouds), and of homogeneous condensates. Section \ref{sec:Holes} is dedicated to the analysis of two more types of solitons: Q-holes and Q-bulges. The analysis is preceded by an extensive motivation to study them, which comes from the analogy between solitons in relativistic field theory and their counterparts in non-relativistic BEC systems and in nonlinear optics. 

In section \ref{sec:Gauge}, we revisit non-topological solitons in theories with additional degrees of freedom. Our goal is to emphasize similarities and dissimilarities between multi-field solutions in those theories and one-field Q-balls. We first consider examples of models with two scalar fields, in which there is a hierarchy of energy scales probed by different field components of a soliton. In this case, the heavy component can be integrated out, resulting in a one-field solution in an effective potential. The effective field theory approach can be used to compute corrections to the solution due to the heavy field. Next, we outline properties of Q-balls in theories with a massless gauge field --- so-called gauged Q-balls. Finally, we touch on the broad subject of boson stars for which the role of an additional field is played by gravity. Section \ref{sec:Summ} contains an outlook on further directions of research in the areas covered by the review.

\textbf{Conventions} Throughout the review, we adopt the natural system of units $c=\hbar=1$. Greek and Latin letters denote space-time and spatial indices correspondingly, and $\dot{\phi}$ stands for the time derivative of $\phi$. The summation is implied over repeated indices. 

\section{Charged scalar condensate and Q-balls}
\label{sec:Gen}

\subsection{Setup}

Consider a theory with a complex scalar field $\phi$ in $d+1$ dimensions with the Lagrangian
\begin{equation}\label{GenLagr}
\mathcal{L}=\eta^{\mu\nu}\partial_\mu\phi^*\partial_\nu\phi-V(\phi^*\phi) \; ,
\end{equation}
where $\eta^{\mu\nu}=\text{diag}(+1,-1,-1,-1)$. Let us find spherically-symmetric configurations $\phi(r,t)$, $r=\sqrt{-\eta^{ij}x_i x_j}$, providing a local extremum of the Hamiltonian
\begin{equation}
H=\int d^dx\:(\dot{\phi}^*\dot{\phi}+\nabla\phi^*\nabla\phi+V(\phi^*\phi))
\end{equation}
among all configurations with a given value of the global $U(1)$-charge 
\begin{equation}
Q=i\int d^dx\:(\dot{\phi}^*\phi-\dot{\phi}\phi^*) \; .
\end{equation}
It amounts to finding an extremum of the functional $H-\omega Q$, where $\omega$ is a Lagrange multiplier. Solving the problem gives \cite{Lee:1991ax}
\begin{equation}\label{GenAnsatz}
\phi(r,t)=f(r)e^{i\omega t} \; ,
\end{equation}
where the function $f(r)$ can be taken real and satisfies the equation of motion
\begin{equation}\label{GenEoM}
\dfrac{d^2f}{dr^2}+\dfrac{d-1}{r}\dfrac{df}{dr}+\omega^2f-f\dfrac{dV(f^2)}{df^2}=0 \; .
\end{equation}
Note that the configuration (\ref{GenAnsatz}) is not static, hence the scaling argument by Derrick \cite{Derrick:1964ww}, which forbids the existence of static solitons in scalar field theories with a non-trivial potential in $d>1$, is evaded.

Eq. (\ref{GenEoM}) is the equation of motion resulting from varying the Lagrangian (\ref{GenLagr}) with respect to $\phi$ and $\phi^*$, followed by applying the ansatz (\ref{GenAnsatz}). In fact, one can as well apply the ansatz (\ref{GenAnsatz}) to the Lagrangian (\ref{GenLagr}) first and then vary with respect to $f$. The principle of symmetric criticality \cite{Palais:1979rca} guarantees that the result will be the same (see also appendix 4 in \cite{Coleman:1975qj}).

A Q-ball is a regular solution of the equation of motion of the form (\ref{GenAnsatz}), with finite charge and energy and with $f$ a monotonic function.\footnote{If the requirement for $f$ to be monotonic is omitted, the profile develops nodes at finite $r$, and one speaks of ``excited states'' of Q-balls (see, e.g., \cite{Mai:2012cx}).} Let us discusss its existence conditions. Without loss of generality we take $V(0)=0$. The qualitative treatment of Q-balls simplifies if one observes that eq. (\ref{GenEoM}) is the Newton's equation describing the motion of a classical particle of unit mass in the potential
\begin{equation}\label{EffPot}
U_\omega(f)=\dfrac{1}{2}(\omega^2f^2-V(f)) \; ,
\end{equation}
and under the influence of friction if $d>1$. In this consideration, the radial coordinate $r$ corresponds to the time parameter, and the scalar field amplitude $f$ corresponds to particle's coordinate. In fact, this mechanical analogy goes beyond the Q-balls, and it reduces the problem of finding different solutions of eq. (\ref{GenEoM}) to the problem of classifying different trajectories of the particle in the potential (\ref{EffPot}). 

\begin{figure}[t]
	\center{\includegraphics[scale=0.55]{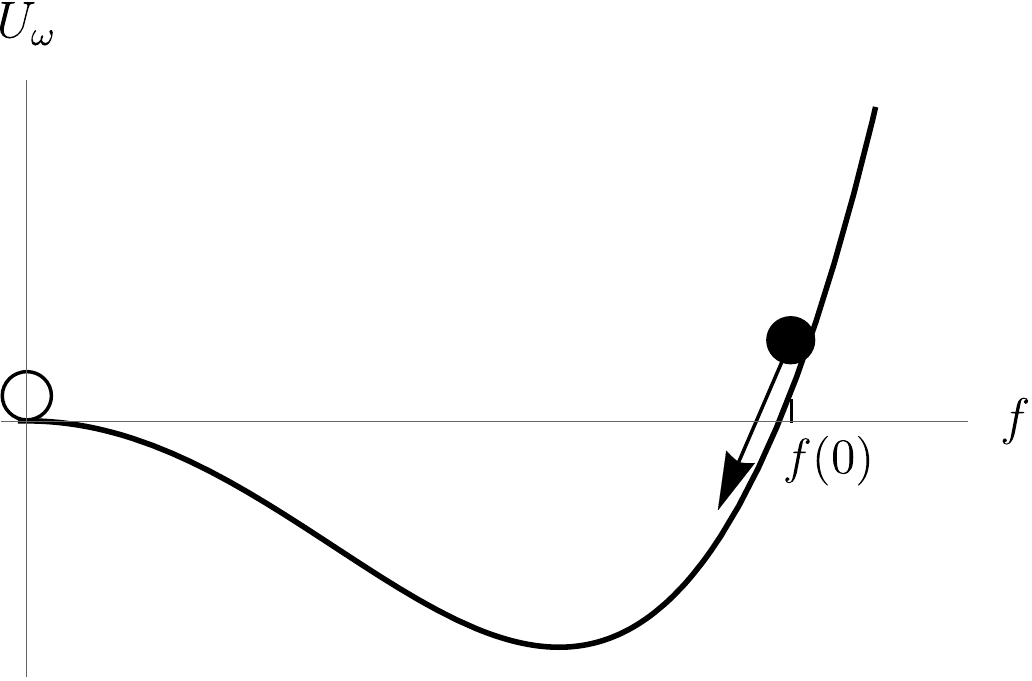}}
	\caption{Schematic form of the potential (\ref{EffPot}) allowing for Q-balls. The trajectory of the particle is shown for the case $d>1$. }
	\label{fig:ball_pot}
\end{figure}

The classical motion corresponding to the Q-balls is shown in Fig. \ref{fig:ball_pot}. The finiteness of the energy implies $f(r)\rightarrow 0$, $r\rightarrow\infty$, hence the particle must finish its motion at the top of $U_\omega(f)$. In particular, one should require $U''_\omega(0)<0$, which can put an upper bound on $|\omega |$. For example, if $V(\phi^*\phi)\sim m^2\phi^*\phi$ near the origin, then $f$ decreases exponentially fast at large $r$, $f\sim e^{-\sqrt{m^2-\omega^2}r}$, and the angular velocity $\omega$ must satisfy the condition $|\omega |<m$. Next, to be able to climb up at the top of $U_\omega(f)$, the particle must start its motion at some point $f(0)$ at which $U'_\omega(f(0))>0$ and $U_\omega(f(0))=0$ if $d=1$, $U_\omega(f(0))>0$ if $d>1$. This can put a non-zero lower bound on $|\omega |$.

\begin{figure*}[t]
	\begin{center}
		\begin{minipage}[h]{0.3\linewidth}
			\center{\includegraphics[scale=0.6]{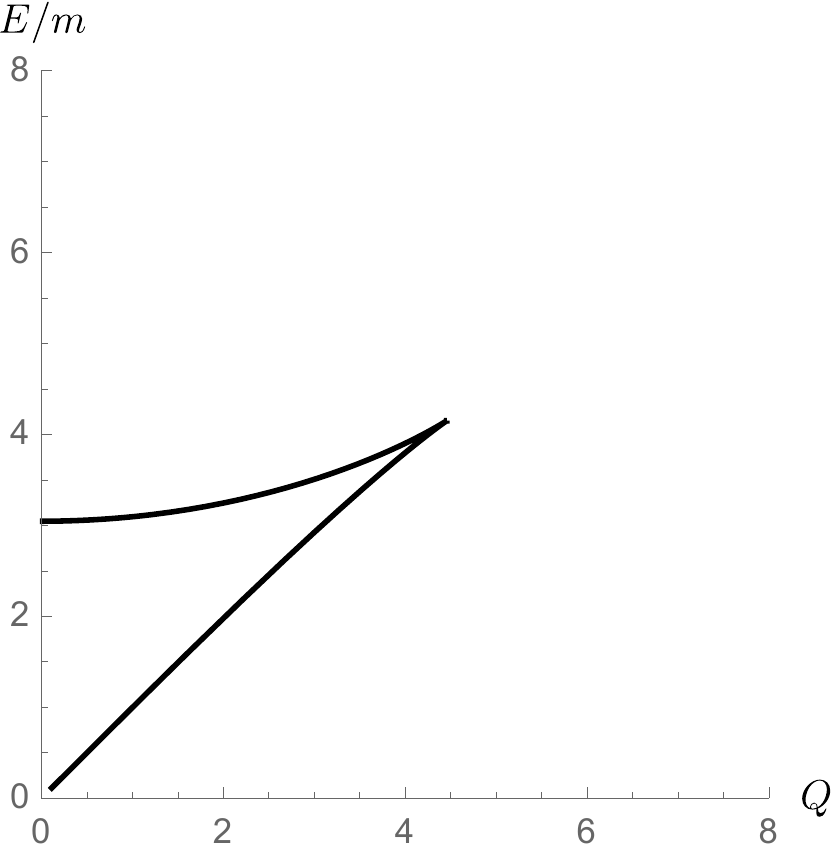} \\ (a)}
		\end{minipage}
		\begin{minipage}[h]{0.3\linewidth}
			\center{\includegraphics[scale=0.6]{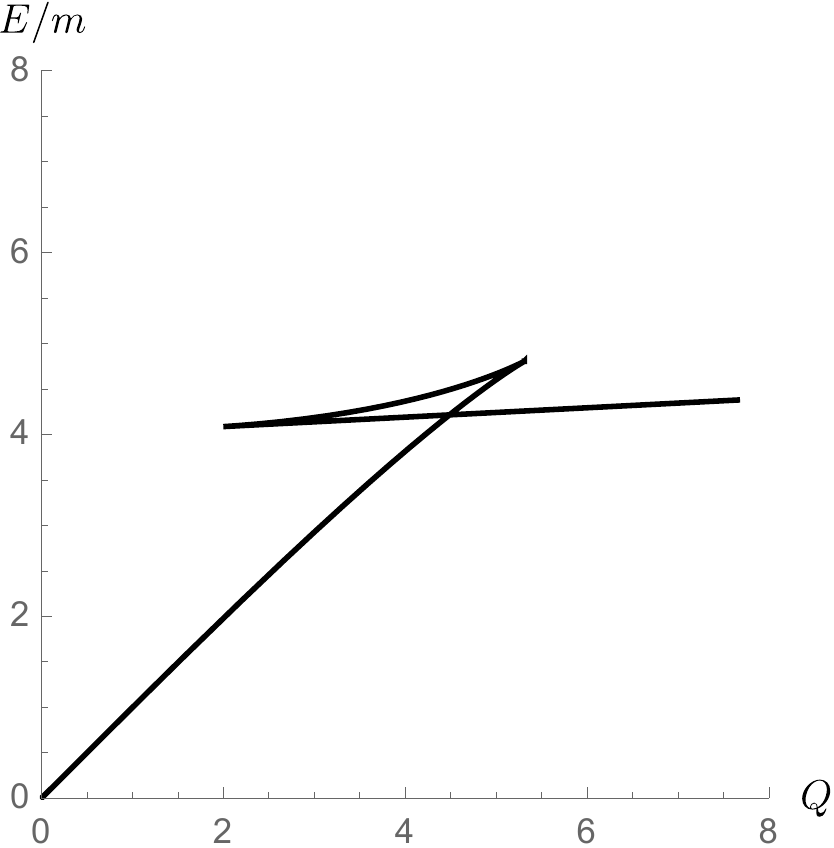} \\ (b)}
		\end{minipage}
		\begin{minipage}[h]{0.3\linewidth}
			\center{\includegraphics[scale=0.6]{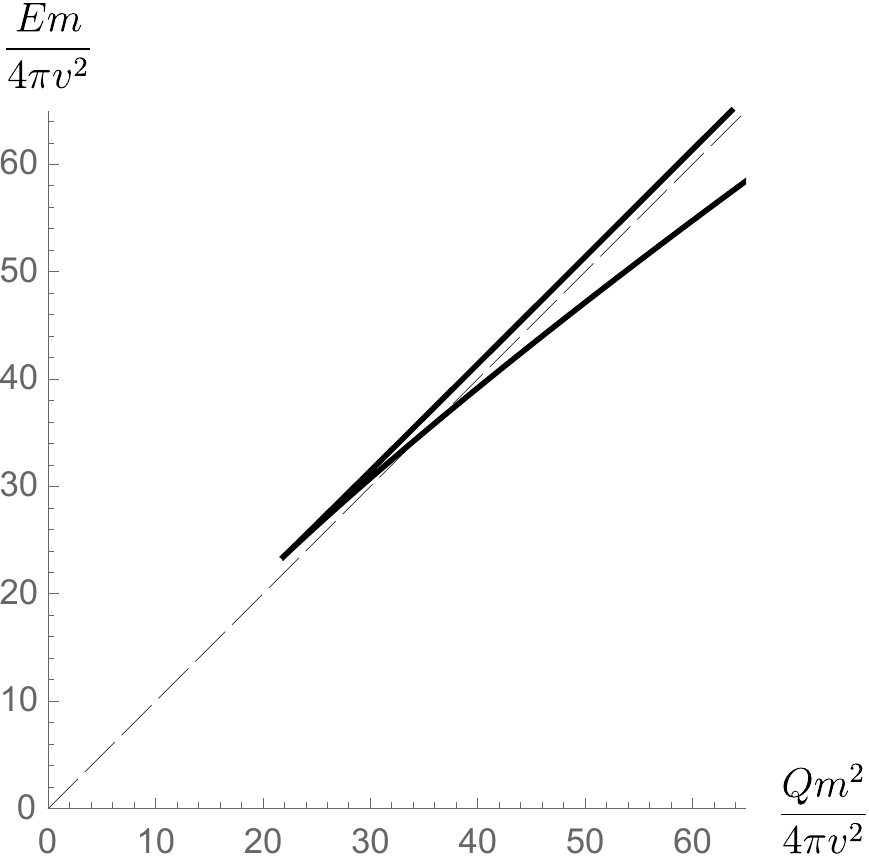} \\ (c)}
		\end{minipage}
		\caption{The energy $E$ of Q-balls plotted against their charge $Q$, in different models. (a): the 1+1-dimensional theory with the polynomial potential of degree six (see eq. (\ref{phi6_Pot})). Q-balls in this theory were studied in \cite{Belova:1994vd}. We take $\alpha=1$ and $\beta=0.1$ so that $\omega_{min}=0$, and there is a static solution. The cusp is on the right. (b): The same theory, but this time $\beta=0.19$ and $\omega_{min}>0$. (c): the 3+1-dimensional theory with the flat piece-wise parabolic potential: $V(\phi\phi^*)=m^2\phi\phi^*$ for $\phi\phi^*<v^2$, and $V(\phi\phi^*)=m^2v^2$ for $\phi\phi^*>v^2$. Q-balls in this potential were considered, e.g., in \cite{Rosen,Theodorakis:2000bz,Gulamov:2013ema}. The cusp is on the left. The dashed line indicates free particle states with $E=mQ$. }
		\label{fig:ball_EQ}
	\end{center}
\end{figure*}

Note that for a given value of $\omega$ in the allowed region, one can launch the particle from different initial positions. Depending on whether the initial position is below or above $f(0)$, the particle will experience either undershoot (that is, it will not reach the maximum at the origin) or overshoot (that is, it will reach the maximum with a non-zero velocity). This makes it easy to search for Q-ball solutions numerically.

Consider now another class of solutions of eq. (\ref{GenEoM}). They are of the form
\begin{equation}\label{Condensate}
f(r)=f_c=\text{Const.}
\end{equation}
The constant solution exists whenever the potential $U_\omega(f)$ develops a local extremum displaced from the origin:
\begin{equation}\label{LocMaxU}
\left.\dfrac{dU_\omega(f)}{df}\right\vert_{f=f_c}=0 \; .
\end{equation}
The solution has finite charge and energy densities,
\begin{equation}\label{RhoCond}
\rho_Q=2\omega f_c^2 \; , ~~~ \rho_E=\omega^2f_c^2+V(f_c^2) \; ,
\end{equation}
but in infinite space its total charge and energy are infinite. Note that the solution does not describe a gas of free particles above the classical vacuum, since, in general, $\rho_E\neq m\rho_Q$. Rather, it represents a homogeneous charged scalar condensate of particles.

If a Q-ball or a condensate solution is classically stable, it describes a time-dependent distribution of matter in the ground state and at zero temperature \cite{Laine:1998rg,Nicolis:2011pv}. This ground state breaks spontaneously time translations associated with the generator $H$ and the internal $U(1)$-symmetry associated with the generator $Q$. However, the linear combination of thereof,
\begin{equation}\label{FreeEnergy}
F := H-\omega Q
\end{equation}
remains unbroken, the situation dubbed in \cite{Nicolis:2011pv} as ``spontaneous symmetry probing''. This observation allows us to identify the angular velocity $\omega$ with the chemical potential of the distribution. As we will see below, the analogy between field theory at zero temperature and statistical physics is fruitful, as it provides an independent look at certain properties of Q-balls and other solitons. As a closing remark, note that finding an extremum of $H$ at a given $Q$ is equivalent to finding an extremum of $F$ at a given $\omega$ \cite{Friedberg:1976me}.

\subsection{Properties}
\label{ssec:Balls_Prop}

In the rest of this section we take $\omega>0$. As was mentioned in Introduction, Q-balls are characterized by their charge and energy:
\begin{equation}\label{GenQE}
\begin{split}
& Q = 2\omega \int d^dx\: f^2 \; , \\
& E = \int d^dx\left(\omega^2f^2+(\nabla f)^2+V(f^2)\right) \; .
\end{split}
\end{equation} 
Taking the derivative with respect to $\omega$ in eqs. (\ref{GenQE}) and using equation of motion (\ref{GenEoM}), one obtains the differential relation between $Q$ and $E$:\footnote{Note that the charge and energy defined by eqs. (\ref{GenQE}) make sense not only for solutions of the equation of motion, but for general configurations of the form (\ref{GenAnsatz}). Hence, they depend on infinite number of parameters. We choose $\omega$ to parameterize one-dimensional sets of Q-balls and other solutions. Then, $\partial Q/\partial\omega$ and $\partial E/\partial\omega$ are understood as directional derivatives of $Q$ and $E$ along a set of solutions. On this set, the relations (\ref{dEdw}) and (\ref{dEdQ}) hold.}
\begin{equation}\label{dEdw}
\dfrac{\partial E}{\partial \omega}=\omega\dfrac{\partial Q}{\partial \omega} \; .
\end{equation}
It means that $Q(\omega)$ and $E(\omega)$ increase or decrease simultaneously and have extrema at the same values of angular velocity, $\omega=\omega_c$. When one draws a parametric plot of $E$ against $Q$, these extrema appear as ``cusp points''. If $\omega\neq\omega_c$, eq. (\ref{dEdw}) is equivalent to
\begin{equation}\label{dEdQ}
\dfrac{\partial E}{\partial Q}=\omega \; .
\end{equation}
Eqs. (\ref{dEdw}) and (\ref{dEdQ}) are the most important relations in the relativistic theory of non-topological solitons. Their range of validity spreads much beyond the Q-balls. In fact, they hold for all types of solitons discussed in this review: homogeneous scalar condensates, nonlinear localized dips and bumps in a condensate, solitons in theories with additional scalar degrees of freedom, with gauge fields and dynamical gravity. 

A cusp point marks a stationary solution with maximal charge and energy (Fig. \ref{fig:ball_EQ}(a)), or minimal charge and energy (Fig. \ref{fig:ball_EQ}(c)), or none of the above (Fig. \ref{fig:ball_EQ}(b)). As we will discuss in section \ref{ssec:Balls_stab}, a cusp point lies at the boundary between the regions of classically stable and classically unstable Q-balls. Note also that in theories with a right cusp (like in Fig. \ref{fig:ball_EQ}(a)), it is possible to have a static Q-ball of zero charge and positive energy. It is natural to associate this Q-ball with a bounce solution mediating transitions between local minima of a potential in quantum mechanics.

Figs. \ref{fig:ball_EQ}(b) and \ref{fig:ball_EQ}(c) demonstrate the existence of solitons with arbitrary large charge and energy. Some of these solutions are absolutely stable. Angular velocities of large stable Q-balls lie close to the lower bound $\omega_{min}$. If this bound is positive, such Q-balls are well described in the thin-wall approximation \cite{Coleman:1985ki}. The wall separates the interior region of a Q-ball, where the field is close to a homogeneous condensate solution, from the exterior region, where the field is close to the classical vacuum. A notion of surface energy is well-defined for Q-balls in the thin-wall regime: 
\begin{equation}
E_{surf}=\int d^dx\:(\nabla f)^2 \; .
\end{equation}
One can also mention the useful relation between the surface energy and the ``free energy'' (\ref{FreeEnergy}) of a Q-ball:
\begin{equation}\label{EwQ}
F=\dfrac{2}{d}\int d^dx\:(\nabla f)^2=\dfrac{2E_{surf}}{d} \; .
\end{equation}
The case $\omega_{min}>0$ is typical for scalar potentials of a polynomial type. In order to allow for Q-balls in a theory with one complex scalar field and a global $U(1)$-symmetry, the (bounded from below) polynomial potential must be at least of degree six \cite{Coleman:1985ki}.\footnote{If one permits unbounded from below potentials, it is enough to have a quartic non-linearity, $V=m^2\phi^*\phi-\lambda(\phi^*\phi)^2$ \cite{Anderson:1970et}. However, the Q-balls in this potential are all unstable.  } Due to their relevance for phenomenology, it is also interesting to consider potentials which become flat at large magnitudes of $f^2$ (see, e.g., \cite{Bezrukov:2007ep,Bezrukov:2009db}). In this case $\omega_{min}=0$, and the thin-wall approximation for large stable Q-balls is not applicable. 

\begin{figure}[t]
	\center{\includegraphics[scale=0.6]{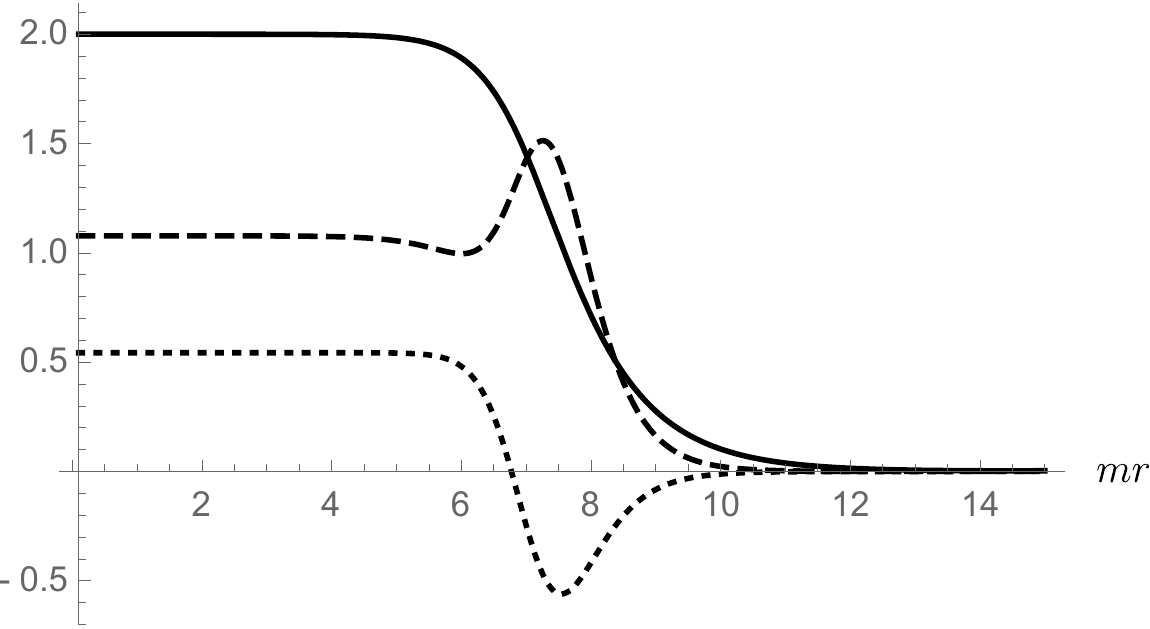}}
	\caption{The distributions of energy (the dashed line) and pressure (the dotted line) of a Q-ball (the solid line) in the thin-wall regime. We consider the 3+1-dimensional theory with the potential (\ref{phi6_Pot}). All quantities are in the units of $m$. The parameters are $\alpha=1$, $\beta=0.2m^{-2}$, $\omega=0.45m$. }
	\label{fig:ball_P_Rho}
\end{figure}

As was mentioned before, stationary configurations of the form (\ref{GenAnsatz}) can be viewed as statistical systems. Hence, it is important to figure out to what extent the similarity between field theory and hydrodynamics goes in the case of nonlinear solutions. To this end, consider the energy-momentum tensor of the theory (\ref{GenLagr}):
\begin{equation}
T_{\mu\nu}=\partial_\mu\phi^*\partial_\nu\phi+\partial_\mu\phi\partial_\nu\phi^*-\eta_{\mu\nu}\mathcal{L} \; .
\end{equation}
At $d>1$, denote
\begin{equation}
T_{00}=\rho_E(r) \; , ~~~ T_{ij}=\left(\dfrac{x_ix_j}{r^2}-\dfrac{1}{d}\delta_{ij}\right)s(r)+\delta_{ij}p(r) \; .
\end{equation}
Then $\rho_E$ is the energy density of the soliton,
\begin{equation}\label{Rho_E}
E=\int d^dx\:\rho_E ,
\end{equation}
$s(r)$ determines the distribution of the shear force,\footnote{To authors' knowledge, the relevance of the shear force to understanding the properties of Q-balls was first pointed out in \cite{Mai:2012yc}. } and $p(r)$ denotes the pressure,
\begin{equation}\label{p(r)}
p(r)=\omega^2f^2-\dfrac{1}{d}f'^2-V(f^2) \; , ~~~ d>1 \; .
\end{equation}
From eqs. (\ref{GenQE}), (\ref{Rho_E}) it follows that for a positive potential $V(f^2)$, the energy density $\rho_E$ is also positive at all $r$. Note that the energy density may not be a monotonically decreasing function of $r$. Next, eqs. (\ref{EffPot}) and (\ref{p(r)}) tell us that in the case $d>1$ there is a region inside a Q-ball where the pressure is positive, but outside the soliton it is always negative. This is illustrated in Fig. \ref{fig:ball_P_Rho} for a 3+1-dimensional Q-ball. We conclude that the thermodynamic equilibrium conditions are not satisfied in the Q-balls. In particular, the ratio $\partial p/\partial\rho_E\equiv\partial p/\partial r\cdot\partial r/\partial\rho_E$, which defines the square of the speed of sound, is negative everywhere but the center of the soliton. In the thin-wall regime both $p$ and $\partial p/\partial\rho_E$ are small and positive inside a Q-ball. As for large stable Q-balls in a flat potential, one can find that for them $\left.\partial p/\partial\rho_E\right\vert_{r=0}\approx 5/3$ when $d=3$. Note finally that for a homogeneous condensate $\rho_E$ is given in eqs. (\ref{RhoCond}) and it is always positive, while $p=2U_\omega(f_c)$, as it follows from eqs. (\ref{EffPot}) and (\ref{p(r)}), and it can be of any sign. 

In the 1+1-dimensional case, there are no off-diagonal components of $T_{ij}$, and the pressure is $p(r)=\omega^2f^2+f'^2-V(f^2)=\text{Const}=0$ on any solution of equation of motion (\ref{GenEoM}).

From the above consideration one concludes that the hydrodynamic description is very limited in its applicability to such systems as Q-balls. It works well in the interior region of large Q-balls in the thin-wall regime and in homogeneous condensates. It breaks down when gradients of fields become important, that is, near the boundary of a soliton. Thus, a general Q-ball is not a ``liquid drop''. The same is true for other non-topological solitons. 

\subsection{Stability}
\label{ssec:Balls_stab}

Several aspects can prevent the absolute stability of a non-topological soliton. First, the soliton can be unstable already at the classical level. This happens when there are small but exponentially growing with time perturbations on top of the solution. Unstable modes can appear in the spectrum of linear perturbations, or at the non-linear level. Linear classical instability means that the (finite-energy) solution does not correspond to a conditional local minimum of the Hamiltonian $H$. Instead, it may correspond to a local maximum or to a saddle point. It turns out that classically unstable Q-balls are saddle points of $H$, because they have exactly one negative mode in their spectra.\footnote{There is no general proof of the uniqueness of a zero mode for classically unstable Q-balls, but it happens to be unique in all explicit calculations, see \cite{Panin:2016ooo} for further discussion.} The homogeneous charged condensates also can be unstable against linear perturbations, and also possess no more than one negative mode.

If a soliton is classically stable, one can ask if the stability is preserved when one takes into account quantum and statistical effects. In particular, one can discuss stability against tunneling into an energetically more favorable state \cite{Lee:1985uv,Levkov:2017paj}. By introducing the interaction of the soliton with other fields, one can ask if it evaporates into quanta of those fields \cite{Cohen:1986ct,Kawasaki:2012gk}, or if thermal fluctuations drive the system out of a local minimum of the energy where the soliton is located \cite{Laine:1998rg}, or if radiative corrections to the scalar potential change it so that instability appears \cite{Kovtun:2016qyj}. In the remainder of this section, we will focus on the linear classical stability of Q-balls and homogeneous condensates.

\subsubsection{Linear classical stability of Q-balls}

Let us first discuss the linear classical stability of Q-balls. Consider the following perturbation of the background solution:
\begin{equation}\label{PertSol}
\phi(\vec{x},t)=f(r)e^{i\omega t}+h(\vec{x},t)e^{i\omega t} \; ,
\end{equation}
where
\begin{equation}\label{PertAnsatz}
h(\vec{x},t)=\sum_{\substack{ l=0 \\ -l\leqslant m\leqslant l }}^\infty \left(h_1^{(l)}(r)e^{i\gamma t}+h_2^{(l)}(r)e^{-i\gamma^* t}\right)Y_{l,m}(\theta,\varphi) \; .
\end{equation}
Here $h_1^{(l)}$ and $h_2^{(l)}$ are real functions, $Y_{l,m}$ are spherical harmonics, and $\gamma$ is, in general, a complex parameter. The perturbed solution (\ref{PertSol}) is substituted into the equation of motion of the field $\phi$, and the latter is linearized with respect to $h_1^{(l)}$, $h_2^{(l)}$. The Q-ball is classically stable, if $\text{Im}\:\gamma=0$ for all solutions of the linearized equation. 

Note that the perturbation ansatz (\ref{PertAnsatz}) does not cover all possible types of perturbations. In particular, the modes corresponding to spontaneously broken $U(1)$ and Lorentz symmetries are not described by this ansatz, see \cite{Smolyakov:2017axd} for further discussion. Note also that, due to the entanglement between $h_1^{(l)}$ and $h_2^{(l)}$, the linearized equation of motion is not reduced to the Schroedinger equation, and the differential operator whose eigenvalues $\gamma$ we are interested in is not hermitian. It is true, however, that some properties of bound states in a Q-ball background can be deduced by using the quantum-mechanical perspective (see, e.g., \cite{Kovtun:2018jae}).

\begin{figure}[t]
	\center{\includegraphics[scale=0.6]{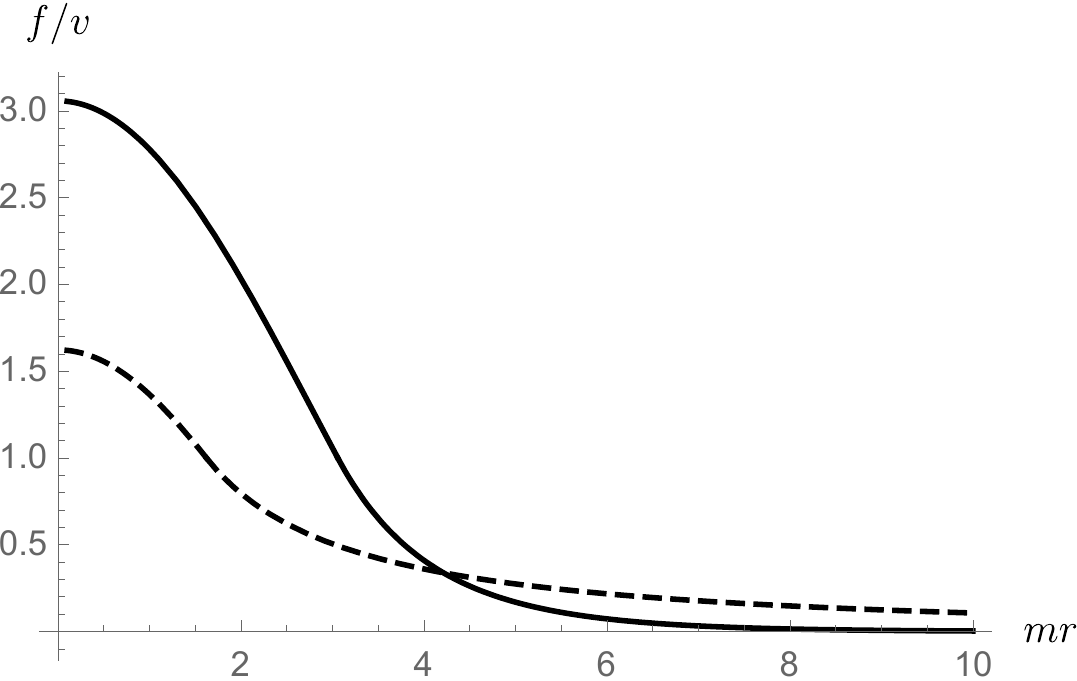}}
	\caption{ A classically unstable Q-ball (or ``Q-cloud'', the dashed line) versus a classically stable Q-ball (the solid line) in the 3+1-dimensional theory with the flat piece-wise parabolic potential (see the caption of Fig. \ref{fig:ball_EQ}). Both solutions have the same charge $Qm^2/(4\pi v^2)\sim 10^2$.  }
	\label{fig:ball_cloud}
\end{figure}

Investigating the conditions under which the linearized operator does not contain exponentially growing with time modes in its spectrum, one arrives at the following sufficient stability condition for Q-balls \cite{Friedberg:1976me,Lee:1991ax,Panin:2016ooo} (see also \cite{Zakharov_2012}):\footnote{It is interesting to note that the condition analogous to (\ref{stab_Q}) was obtained earlier for solutions of the Nonlinear Schroedinger Equation \cite{vakhitov,Kolokolov1973}.}
\begin{equation}\label{stab_Q}
\dfrac{\partial Q}{\partial\omega}<0 \; ,
\end{equation}
or, in view of eq. (\ref{dEdw}),
\begin{equation}\label{stab_Q_2}
\dfrac{\partial^2E}{\partial Q^2}<0 \; .
\end{equation}
Studies revealed that eq. (\ref{stab_Q}) is also necessary for the classical stability of Q-balls \cite{Panin:2016ooo}. We see that the regions of stability and instability are separated by the cusp points at which $\partial Q/\partial\omega=\partial E/\partial\omega=0$. According to eq. (\ref{stab_Q_2}), the lower branches of Q-balls in Fig. \ref{fig:ball_EQ} are stable, and the upper branches are unstable. Large unstable Q-balls have more dispersed profiles than their stable counterparts of the same charge, as is illustrated in Fig. \ref{fig:ball_cloud}. Because of this, they are also referred to as Q-clouds \cite{Alford:1987vs}.  

As was discussed in section \ref{ssec:Balls_Prop}, the angular velocity $\omega$ of a Q-ball can be associated with the chemical potential of the system. On the other hand, the charge $Q$ of the soliton is proportional to the number of particles $N$ constituting it. Eq. (\ref{stab_Q}) can then be interpreted as follows. If we want to remove one particle from the Q-ball, we should apply a work $-\partial\omega/\partial Q\cdot\Delta Q$ with $\Delta Q\sim \Delta N=1$. If the work is negative, the Q-ball favors the process of disintegration into individual quanta; otherwise it is stable against this process. The classical stability condition is, therefore, reproduced by the statistical argument.

However, the statistical argument has its limits of applicability. It breaks down if the classical solution whose stability we investigate cannot be isolated from spatial infinity at which a test particle is located. This is the case of extended solitons such as elongated, vortex-like axially-symmetric Q-tubes. And, indeed, it is known that for Q-tubes the condition (\ref{stab_Q}) does not work \cite{Nugaev:2014iva}. This is also the case when the theory (\ref{GenLagr}) is supplemented with massless fields generating long-ranged forces. For example, adding the gauge spin-1 field gives rise to so-called gauged Q-balls \cite{Rosen2}, to which the condition (\ref{stab_Q}) is also inapplicable \cite{Panin:2016ooo}, see section \ref{ssec:Gauged} for more detail.

Let us go back to the discussion of the spectrum of linear perturbations of the form (\ref{PertAnsatz}) above a Q-ball. Studies revealed that classically unstable Q-balls possess a single spherically-symmetric decay mode \cite{Lee:1991ax,Smolyakov:2017axd}. For this mode Re$\:\gamma=0$, and there are no more spherically-symmetric solutions of the linearized equation of motion of the form (\ref{PertAnsatz}). An unstable Q-ball is, therefore, a saddle point of the Hamiltonian. For stable Q-balls we have Im$\:\gamma=0$, and the linearized equation gives rise to a spectrum of vibrational modes. In the limit of large $Q$, their number is proportional to the volume of the soliton $\sim Q^3$ \cite{Kovtun:2018jae}. Note finally that when one crosses the cusp point $\omega=\omega_c$, the decay mode of unstable Q-balls (with $\gamma^2<0$) converts into a vibrational mode of stable Q-balls (with $\gamma^2>0$). The solution with $\omega=\omega_c$ has, therefore, an additional zero mode in its spectrum.

\subsubsection{Linear classical stability of homogeneous condensates}

The necessary and sufficient condition for the absence of growing modes in the perturbations of the form (\ref{PertAnsatz}) above a homogeneous solution (\ref{Condensate}) is written as follows (see, e.g., \cite{Nugaev:2015rna}): 
\begin{equation}\label{stab_c}
\left.\dfrac{\partial^2V(z)}{\partial z^2}\right\vert_{f=f_c}>0 \; , ~~~ z=\phi^*\phi\; .
\end{equation}
First, we note that if the potential $V$ is bounded from below, it must satisfy the condition (\ref{stab_c}) at some magnitude $f_c$ in order to allow for Q-balls \cite{Coleman:1985ki}. Next, it is interesting to compare this condition with the criterion (\ref{stab_Q}) of classical stability of Q-balls. Since both charge and energy of a condensate diverge in infinite space, we impose the periodic boundary conditions with a large period $L$.\footnote{It is possible to make $L$ much larger than the size of the Q-ball with the charge equal to $Q_c$. } Then,
\begin{equation}
Q_c=L^d \rho_Q \; , ~~~ E_c=L^d\rho_E \; ,
\end{equation}
where $\rho_Q$, $\rho_E$ are given in eqs. (\ref{RhoCond}). Eq. (\ref{stab_c}) becomes
\begin{equation}\label{stab_cond}
\left.\dfrac{\partial Q_c}{\partial \omega}\right\vert_{\omega=\omega_c}>0 \; ,
\end{equation}
and it is opposite to the criterion (\ref{stab_Q}). 

The difference in the stability conditions implies a difference in the properties of the corresponding background solutions. Indeed, one can check that the method of deriving eq. (\ref{stab_Q}) does not work for homogeneous solutions \cite{Panin:2016ooo}. Furthermore, the statistical argument also fails, since it is impossible to isolate a particle from the system occupying all available volume. Note, however, that eqs. (\ref{stab_Q}) and (\ref{stab_cond}) do not exclude each other, and it is not difficult to construct a scalar potential which admits both classically stable Q-balls and classically stable condensates \cite{Nugaev:2015rna}.

Let us make a comment regarding the above observation. The common mechanisms of formation of Q-balls in the Early Universe rely on the existence of classically unstable homogeneous condensate decaying into long-lived localized nonlinear configurations \cite{Kolb:1993zz,Kusenko:1997hj,Kusenko:1997si,Khlebnikov:1999qy,Kasuya:2000wx,Amin:2010xe,Krylov:2013qe} (see also \cite{zakharov1968}). These mechanisms are not applicable to theories with a stable condensate, and to achieve the fragmentation of the homogeneous background into Q-balls in this case is a challenge.\footnote{The similar question is discussed in \cite{Lee:1985uv}.} A possible formation mechanism includes quantum transitions between homogeneous and non-homogeneous configurations \cite{Nugaev:2015rna,Levkov:2017paj}. More precisely, the condensate can tunnel into an energetically more favorable Q-ball of (nearly) the same charge. Moreover, the path of the bounce solution responsible for the tunneling goes through the saddle point corresponding to a Q-cloud. Thus, the Q-cloud is a sphaleron, that is, a configuration representing the top of the barrier separating classically stable solutions. The role played by Q-clouds in quantum transitions between Q-balls and condensates was hypothesized in \cite{Nugaev:2015rna} and confirmed in \cite{Levkov:2017paj}.

Note finally that the decay mode of a condensate, if exists, is unique, and the corresponding eigenvalue $\gamma$ is purely imaginary, as in the case of Q-balls.

\section{Nonlinear dips and rises in charged scalar condensate}
\label{sec:Holes}

\subsection{Motivation}
\label{ssec:Motivation}

\subsubsection{Nonlinear optics}

As was mentioned in Introduction, solitons arise in many areas of modern physics beyond relativistic field theory. Nonlinear optics is another branch of physics, perhaps, the most convenient for experiment, where they are studied \cite{kivshar2003optical,akhmediev2005dissipative}. In particular, propagation of quasi-one dimensional nonlinear pulses along a single-mode optical fibers has been attracting significant attention since 1970s (for a review, see \cite{KUMAR199063}). An optical soliton in this case is an envelope $u$ of a complex amplitude of an electromagnetic wave running through the fiber. Save inevitable fiber losses, the balance between the dispersive broadening of a pulse and the nonlinear change in the refractive index of the fiber can stabilize the envelope, resulting in a distortionless propagation of the soliton. 

A distortionless transmission of optical pulses in fibers was first discussed by Hasegawa and Tappert in 1973 \cite{Hasegawa1,Hasegawa2}, and was first observed experimentally in 1980 \cite{Mollenauer:1980zz}. Today, optical solitons are in focus of the research in view of their potential applications to long-distance fiber communications (in commercial use since 2002), optical switching, logical gates etc \cite{Hasegawa_3,Blair2002,mitschke2012recent}.

The basic equation in the theory of optical solitons is the Nonlinear Schroedinger Equation (NSE). In an appropriate system of coordinates, this equation is
\begin{equation}\label{NSE_NLO}
iu_z-\sigma u_{tt}+2|u|^2u=0 \; .
\end{equation}
Here $z$ is a distance along the fiber, $t$ is a retarded time measured in a reference frame moving along the fiber at the group velocity, and $\sigma=\pm 1$. Depending on the sign of $\sigma$, two regimes are realized. The case $\sigma=-1$ corresponds to the negative group velocity dispersion. Eq. (\ref{NSE_NLO}) then possesses bright soliton solutions. For them, $u$ has the form of a bump which decreases rapidly with $t$. Eq. (\ref{NSE_NLO}) also has a constant wave solution $|u|=\:$Const. analogous to the charged scalar condensate, and this solution is classically unstable for $\sigma=-1$. 

The case $\sigma=1$ corresponds to the positive group velocity dispersion. In this regime bright solitons are forbidden. The constant wave solution still exists and is classically stable. Apart from it, eq. (\ref{NSE_NLO}) admits solutions of the form of localized nonlinear dips in the homogeneous background $|u|=$Const. These were dubbed dark solitons \cite{Kivshar}.\footnote{The fundamental solution of the NSE describing the dark soliton was first obtained in \cite{Zakharov_Shabat}.} While for bright solitons there is no shift in the phase of the complex envelope, a dark soliton experiences a phase jump across its center. Dark solitons were first observed in 1981 on top of a broad bright pulse imitating the constant wave background \cite{Weiner}.

\subsubsection{Bose-Einstein condensate}

Bose-Einstein condensation \cite{PitaevskiiBook} in dilute atomic gases \cite{pethick_smith_2008} was first detected in 1995 in experiments on vapors of Rubidium, Sodium and Lithium; for a review see \cite{Dalfovo:1999zz}. Nowadays, the study of dilute quantum gases is a major part of molecular, atomic and condensed matter physics. The significance of the Bose-Einstein Condensate (BEC)\footnote{The Bose-Einstein condensate is not to be confused with the scalar condensate discussed in section \ref{sec:Gen}. In our notations, the latter is relativistic and homogeneous, while the former is non-relativistic and, in general, non-homogeneous. } for fundamental physics is revealed, e.g., in studies of sonic black holes \cite{Garay:2000jj,Lahav:2009wx}. It is also relevant for astrophysics and cosmology, e.g., as a dark matter candidate \cite{Suarez:2013iw}.

The existence of BEC relies on external forces confining an atomic gas in a trap, usually, a harmonic one. The external potential is provided by a magnetic field. Typically, BEC is a dilute ground state of the system, which is non-homogeneous both in momentum and coordinate spaces. The mean free path of an atom inside a BEC can be orders of magnitudes larger than the range of inter-atomic forces. Despite this, the role of two-body interactions is crucial in the formation of the condensate.

\begin{figure*}[t]
	\begin{center}
		\begin{minipage}[h]{0.45\linewidth}
			\center{\includegraphics[scale=0.55]{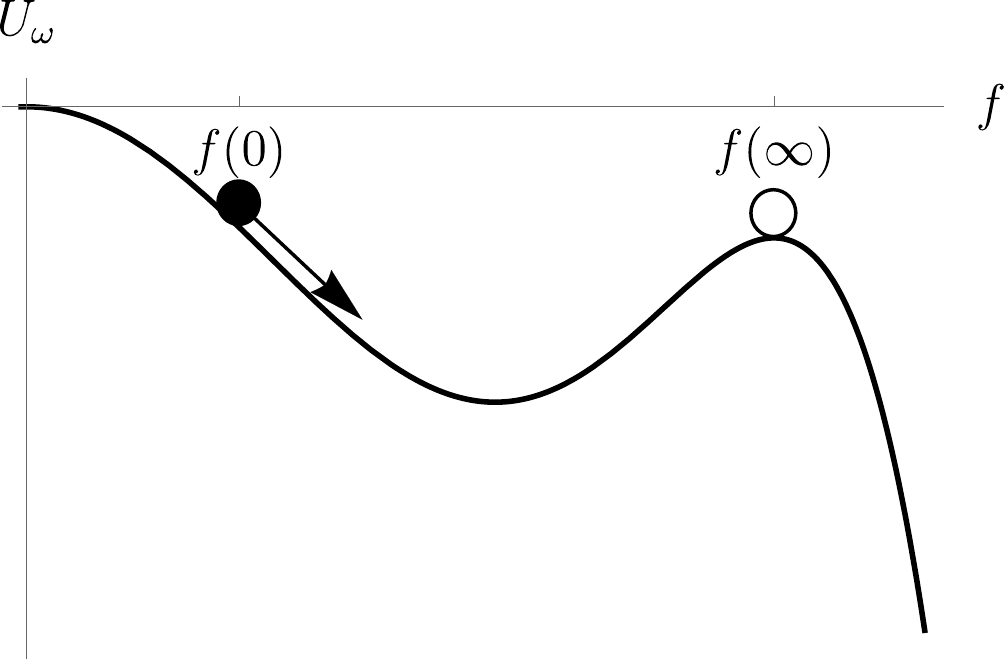} \\ (a)}
		\end{minipage}
		\begin{minipage}[h]{0.45\linewidth}
			\center{\includegraphics[scale=0.55]{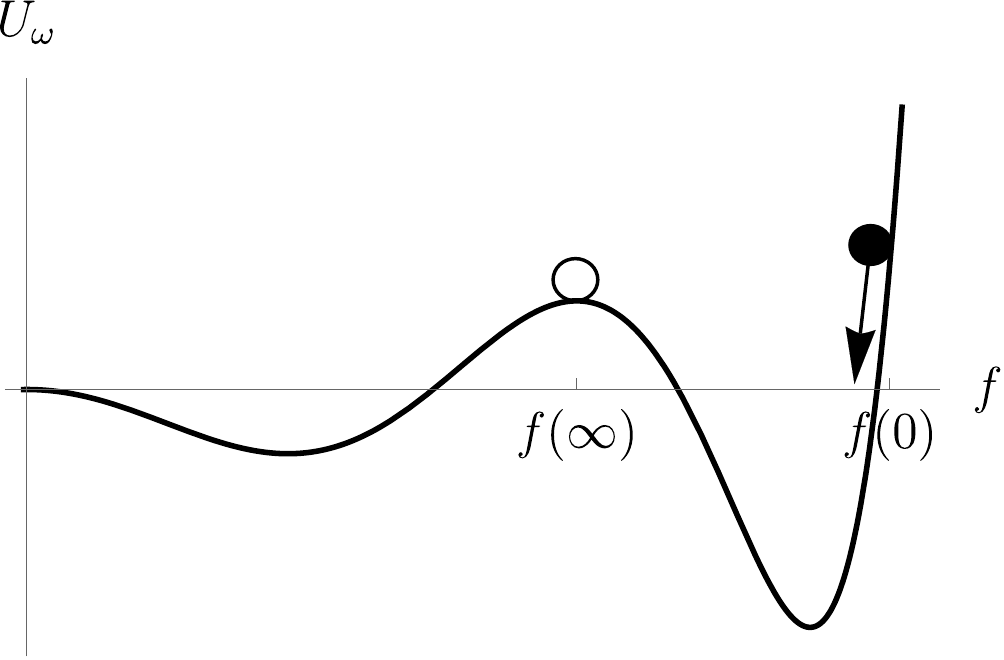} \\ (b)}
		\end{minipage}
		\caption{Schematic forms of the potential (\ref{EffPot}) allowing for Q-holes (a) and Q-bulges (b). The trajectory of the particle is shown for the case $d>1$.}
		\label{fig:holebulge_pot}
	\end{center}
\end{figure*}

The condensate and its excitations in the mean-field approximation are described by the Gross-Pitaevskii equation. The latter has the form of the NSE. For example, consider a system in a spherically-symmetric trapping potential $V(r)$. Then, in appropriate units (cf. eq. (\ref{NSE_NLO}))
\begin{equation}\label{NSE_BEC}
i\psi_t+\Delta_r\psi-g|\psi|^2\psi+V(r)\psi=0 \; .
\end{equation}
Here $\psi$ is a wavefunction of the system, $t$ and $r$ are time and space radial coordinates accordingly, and $g$ determines the strength of the inter-atomic force. Apart from the ground-state BEC, eq. (\ref{NSE_BEC}) may admit nonlinear solitons analogous to optical pulses discussed above. If the inter-atomic force is attractive, $g>0$, then there are bright solitons or vortices. These were first observed in a two-component BEC in 1999 \cite{Matthews:1999zz}. If the force is repulsive, $g<0$, then there are dark solitons. They were also discovered in experiment \cite{Burger:1999zz}. 

Note that the solution of the Gross-Pitaevskii equation (\ref{NSE_BEC}) minimizes the energy functional
\begin{equation}
E[\psi]=\int dr\:\left(|\nabla\psi|^2+V(r)|\psi|^2+\dfrac{g}{2}|\psi|^4\right) \;
\end{equation}
for a fixed number of particles $N=\int dr\:|\psi|^2$. Thus, the non-relativistic theory describing the behaviour of BEC is invariant under constant shifts of the phase of the wavefunction $\psi$, much like the relativistic theory (\ref{GenLagr}) describing Q-balls is invariant under constant shifts of the phase of the complex field $\phi$.

\subsubsection{Relativistic field theory}

As we have seen, choosing the sign of the group velocity dispersion or the sign of the inter-atomic force results in the appearance of bright or dark solitons. Bright solitons have obvious analogs in relativistic field theory. One can think of Q-balls as their distant relatives. For example, in \cite{Enqvist:2003zb} the analogy between Q-balls and solitons in BEC was elaborated, and in \cite{Bunkov:2007fe} the actual formation of Q-balls in BEC was discussed.

Classical solutions of relativistic field equations that would correspond to dark solitons are also easy to find. Consider, for example, the 1+1-dimensional theory of the real scalar field $\varphi$ with the mexican-hat potential, $V(\varphi)=m^2\varphi^2-\lambda\varphi^4$, $\lambda>0$. It yields the usual kink solution, $\varphi\sim\tanh (mx/\sqrt{2})$ \cite{Dashen:1974cj,Polyakov:1974ek,Goldstone:1974gf}. Let us promote $\varphi$ to the complex field $\phi$ with replacing $\varphi^2\rightarrow\phi^*\phi$ in the Lagrangian to make the latter $U(1)$-invariant. Then, the theory admits stationary solutions of the form (\ref{GenAnsatz}). At $\omega=0$ the original kink of zero charge is reproduced. At $\omega>0$ we have a family of complex kinks with the charge density
\begin{equation}
\rho_Q=2\omega f^2 \; .
\end{equation}
Clearly, $\rho_Q$ has a dip at the center of the configuration. Away from the center, the charged scalar condensate is restored.\footnote{Such complex kinks can be identified with sphalerons in the abelian Higgs model, see chapter 11 of \cite{Manton:2004tk}. } Note that the asymptotics of $\phi$ at $x\rightarrow\pm\infty$ have different phases, in full analogy with the dark soliton solution of the NSE.  

It is natural to ask if there are analogs of dark solitons among non-topological configurations similar to Q-balls. The answer is affirmative \cite{Nugaev:2016wyt}. In theories with a complex scalar field these configurations describe local nonlinear dips (``Q-holes'') in a homogeneous condensate. Unlike dark solitons, the field topology of a Q-hole at spatial infinity is trivial, and, what is more, Q-holes can exist in the same theories as Q-balls. Apart from Q-holes, there may also exist local nonlinear bumps (``Q-bulges'') on top of a condensate background. Both Q-holes and Q-bulges require the surrounding condensate to be classically stable. 

The rest of this section is dedicated to the detailed discussion of the two types of solitons. In section \ref{ssec:Holes_QQ}, we discuss existence conditions of Q-holes and Q-bulges and their main features. We also touch upon the question of classical stability of Q-holes and Q-bulges. The exposition there is based on \cite{Nugaev:2016wyt}. In section \ref{ssec:Holes_ex}, an explicit example is considered, based on a 1+1-dimensional model with a polynomial potential.

\subsection{Q-holes and Q-bulges}
\label{ssec:Holes_QQ}

Consider a theory with a complex scalar field in $d+1$ dimensions with the Lagrangian (\ref{GenLagr}). Let us apply the ansatz (\ref{GenAnsatz}) for the scalar field. In section \ref{sec:Gen} we employed the analogy between soliton profiles and motions of a classical particle in the potential (\ref{EffPot}) to find Q-ball solutions. We will now use this analogy to reveal Q-holes and Q-bulges.

Let the potential $U_\omega$ develop a local maximum displaced from the origin at some $\omega$, see Fig. \ref{fig:holebulge_pot}(a). Then there is a trajectory that starts at some $f(0)>0$ and moves towards that maximum, reaching it at $r=\infty$, $f(\infty)>f(0)$. This trajectory corresponds to the soliton profile $f$ that increases monotonically from its center and approaches some constant magnitude at spatial infinity. Associated with the local maximum of $U_\omega$ is the homogeneous condensate solution eq. (\ref{LocMaxU}). Therefore, the soliton represents the local nonlinear dip in the scalar condensate, hence the name Q-hole.

Let now the potential $U_\omega$ grow at some $f$ above the position of the local maximum, see Fig. \ref{fig:holebulge_pot}(b). In this case, there is a trajectory that starts at some $f(0)$ above the maximum and moves towards the maximum, reaching it at $r=\infty$, $f(\infty)<f(0)$. The trajectory corresponds to a solution which decreases monotonically with $r$ and approaches the condensate at spatial infinity. Therefore, it represents the local nonlinear bump in the scalar condensate, hence the name Q-bulge.

\begin{figure*}
	\begin{center}
		\begin{minipage}[h]{0.32\linewidth}
			\center{\includegraphics[scale=0.67]{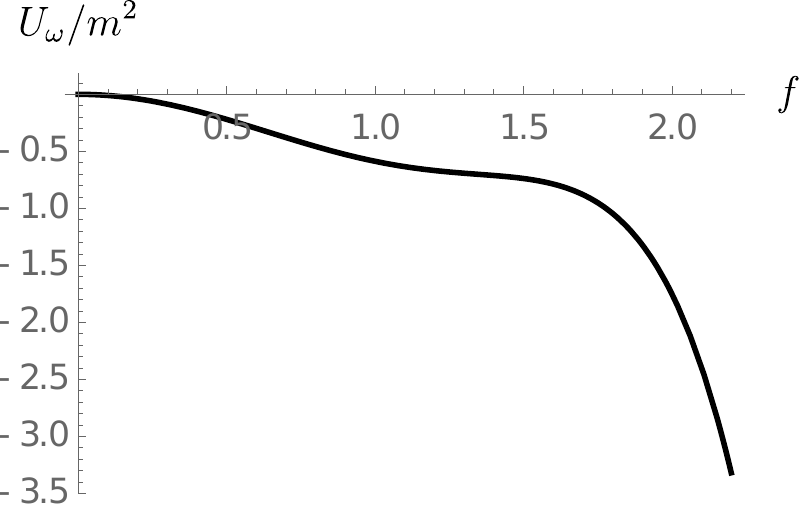} \\ (a)}
		\end{minipage}
		\begin{minipage}[h]{0.32\linewidth}
			\center{\includegraphics[scale=0.67]{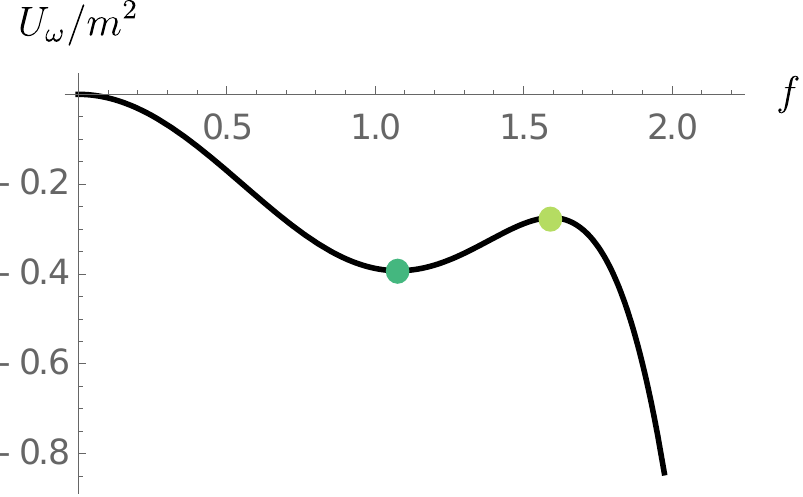} \\ (b)}
		\end{minipage}		
		\begin{minipage}[h]{0.32\linewidth}
			\center{\includegraphics[scale=0.67]{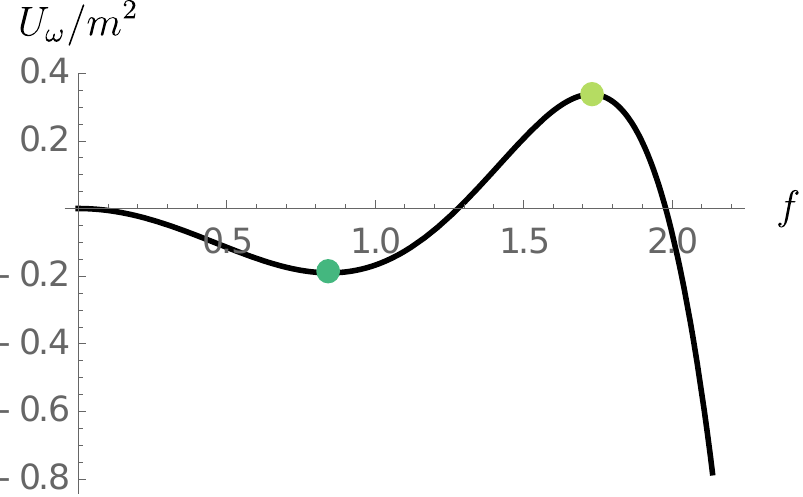} \\ (c)}
		\end{minipage}		
		\caption{The potential (\ref{EffPot}) with $V$ given in eq. (\ref{phi6_Pot}), at different $\omega$. The colored dots denote the condensate solutions. }
		\label{fig:holes_pot}
	\end{center}
\end{figure*}

\begin{figure}[t]
	\center{\includegraphics[scale=0.6]{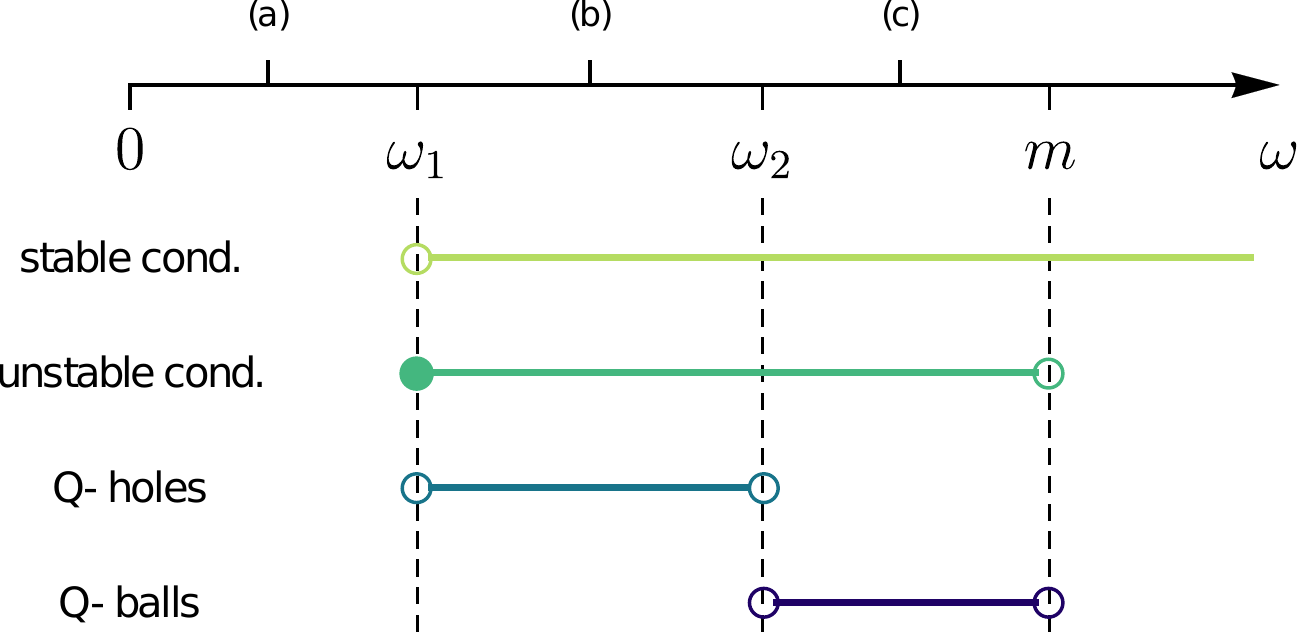}}
	\caption{ The spectrum of classical solutions in the theory with the potential (\ref{phi6_Pot}). The ticks (a), (b) and (c) refer to the potential $U_\omega$ shown in Fig. (\ref{fig:holes_pot})(a,b,c) correspondingly. }
	\label{fig:holes_ranges}
\end{figure}

The crucial feature of Q-holes and Q-bulges is that at large distances they do not approach the classical vacuum of the theory. Instead, they approach a certain condensate solution, and the angular velocity of the soliton is synchronized with the angular velocity of the background condensate.

The charge $Q$ and the energy $E$ of Q-holes and Q-bulges are defined in the usual way, eqs. (\ref{GenQE}). In infinite space they both diverge. Hence, it makes sense to consider the values of $Q$ and $E$ relative to the charge $Q_c$ and the energy $E_c$ of the background condensate:
\begin{equation}
Q_{rel}=Q-Q_c \; , ~~~ E_{rel}=E-E_c \; .
\end{equation}
Then, both $Q_{rel}$ and $E_{rel}$ are finite and satisfy the relations (\ref{dEdw}), (\ref{EwQ}). A Q-hole (Q-bulge) $\phi(r,t)$ is a configuration providing a finite-value extremum of the functional $E_{rel}=E_{rel}(\omega,f_c)$ at a given value of $Q_{rel}$ and having $f_c=f_c(\omega)$ and $\omega$ (the magnitude and the angular velocity of the condensate) fixed. Indeed, the variational problem
\begin{equation}
\delta(E_{rel}(\omega,f_c)-\Omega Q_{rel})=0
\end{equation}
gives $\phi=f(r)e^{i\Omega t}$ with $f$ satisfying eq. (\ref{GenEoM}), and the requirement $|E_{rel}|<\infty$ gives $\Omega=\omega$, $f\rightarrow f_c$, $r\rightarrow\infty$.\footnote{In finite volume the requirement $|E_{rel}|<\infty$ is unnecessary, and the angular velocity of a Q-hole (Q-bulge) does not, in general, coincide with the angular velocity of a condensate. }

The relative charge of Q-bulges is always positive and so is their relative energy, see eq. (\ref{EwQ}). For Q-holes $Q_{rel}<0$, and $E_{rel}$ is not positive-definite. Hence, it is possible for a condensate with a hole to be more energetically favorable state than the condensate of the same charge and without a hole. This happens when the energy gain due to the energy density recess inside a Q-hole exceeds the energy needed to build the walls of the hole. Whether $E_{rel}$ can be negative depends on the model under investigation. For example, this is the case in the theory with the piece-wise parabolic potential studied in \cite{Nugaev:2016wyt}. This is not the case in the theory with the polynomial potential considered below. 

Q-holes and Q-bulges can exist only on top of a classically stable condensate solution. Hence, even if a condensate with a hole is energetically more favorable, the hole cannot appear as a result of linear classical instability of the background. Instead, one can hypothesize that Q-holes are an intermediate product of quantum decay of homogeneous configurations.

To allow for Q-holes, the second maximum of the potential $U_\omega$ must be lower than the maximum at the origin, as demonstrated in Fig. \ref{fig:holes_pot}(b). On contrary, to allow for Q-balls, the second maximum must be higher than the first one, as shown in Fig. \ref{fig:holes_pot}(c). But for a given potential $V$, the value of $U_\omega$ at the second maximum is controlled by $\omega$. The allowed region of angular velocities for Q-balls is $\omega_{min}\equiv\omega_2<|\omega|<m$, where $\omega_2$ is the value at which the maxima degenerate (or zero), and the upper bound was discussed in section \ref{sec:Gen}. If $\omega_2>0$, then the theory admits Q-holes and the valules of $\omega$ at which they exist are complementary to those of Q-balls: $\omega_1<|\omega|<\omega_2$, where $\omega_1$ is the value at which the second maximum disappears (or zero). We conclude that Q-holes are not exotic objects in theories with peculiar potentials; instead, they are very typical and can exist in the same models as Q-balls, just at lower angular velocities. We will see this in an explicit example below.

There is no definite conclusion about classical (in)stability of Q-holes and Q-bulges. Note that the criterion (\ref{stab_Q}) is not applicable to configurations with non-vacuum asymptotics at spatial infinity \cite{Panin:2016ooo}. An argument in favour of their instability was provided in \cite{Nugaev:2016wyt}, and it is based on considering first the system in a finite box of size $L$, and then taking the limit $L\rightarrow\infty$. Let us repeat the argument here for Q-holes; for Q-bulges it goes identically. For simplicity, consider the 1+1-dimensional case, and impose the periodic boundary conditions $f(-L/2)=f(L/2)$, $f'(-L/2)=f'(L/2)$. Nothing precludes us to apply the condition (\ref{stab_Q}) to finite-size configurations \cite{Panin:2016ooo}, and a Q-hole in a box is classically unstable if
\begin{equation}\label{stab_hole}
\dfrac{\partial Q_L}{\partial\omega}>0 \; ,
\end{equation}
where $Q_L$ denotes the finite charge of such a Q-hole. Let us now investigate the regime when $L$ is much larger than the characteristic size of the soliton. Then, there exists the homogeneous condensate of magnitude $f_{c,L}$ and charge $Q_{c,L}$ such that

\begin{equation}
\begin{split}\label{CondHoles}
& f(\pm L/2)-f_{c,L}\rightarrow 0 \; ,\\ & \frac{Q_L-Q_{c,L}}{Q_{c,L}}\rightarrow 0 \; , \\
& \left(\frac{\partial Q_L}{\partial\omega}-\frac{\partial Q_{c,L}}{\partial\omega}\right)\frac{1}{Q_{c,L}}\rightarrow 0 \; , ~~~ L\rightarrow\infty \; .
\end{split}
\end{equation}
Since a condensate solution a Q-hole approaches at large distances is required to be classically stable, in view of eq. (\ref{stab_cond}) it is reasonable to expect that at large enough $L$
\begin{equation}
\dfrac{\partial Q_{c,L}}{\partial\omega}>0 \; .
\end{equation}
This and the last of eqs. (\ref{CondHoles}) then imply that eq. (\ref{stab_hole}) holds at large enough $L$. Thus, there is a decay mode associated with the Q-hole confined in a finite volume. The length scale of the mode is determined by the length scale of the Q-hole and does not depend on $L$ at large $L$. Hence, the mode survives in the limit $L\rightarrow\infty$ and renders the Q-hole in infinite space unstable.

\begin{figure}[b]
	\center{\includegraphics[scale=0.7]{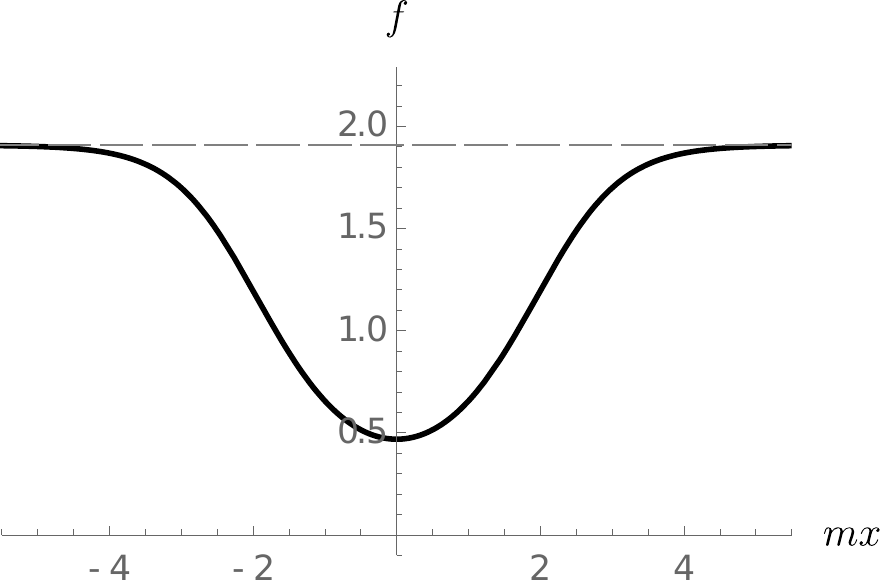}}
	\caption{ The Q-hole in the potential (\ref{phi6_Pot}). Here $\alpha=m$, $\beta=0.2m^2$, and $\omega=0.1m$. The dashed line denotes the magnitude of the background condensate. }
	\label{fig:holes_profile}
\end{figure}

\begin{figure*}
	\begin{center}
		\begin{minipage}[h]{0.49\linewidth}
			\center{\includegraphics[scale=0.67]{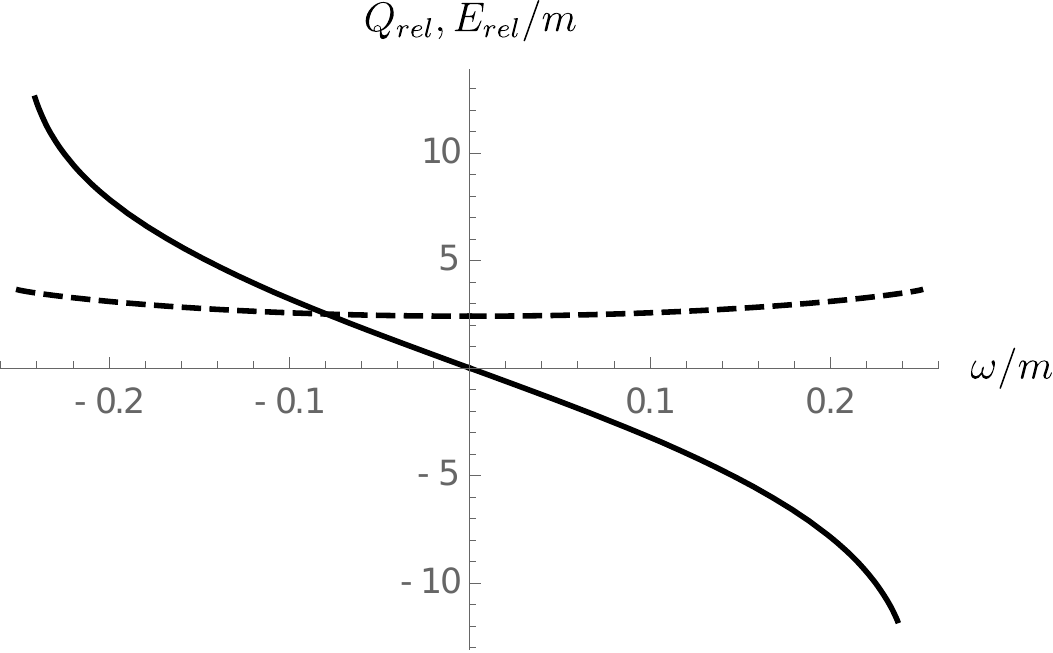} \\ (a)}
		\end{minipage}
		\begin{minipage}[h]{0.49\linewidth}
			\center{\includegraphics[scale=0.67]{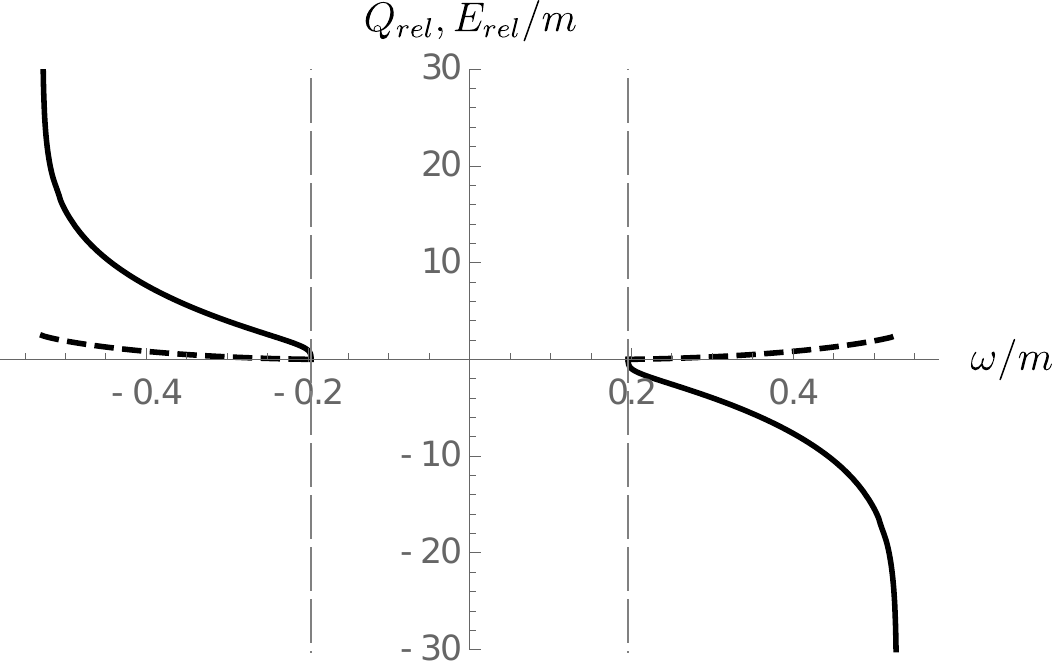} \\ (b)}
		\end{minipage}				
		\caption{The relative charge (the solid lines) and energy (the dashed lines) of Q-holes in the potential (\ref{phi6_Pot}). (a): $\alpha=m$, $\beta=0.12m^2$. According to eq. (\ref{w1}), the lower bound is $\omega_1=0$. (b): $\alpha=m$, $\beta=0.26m^2$. Here $\omega_1=0.2m$, the bound is denoted by the vertical dashed lines. }
		\label{fig:holes_QE}
	\end{center}
\end{figure*}

\subsection{An example}
\label{ssec:Holes_ex}

Consider a 1+1-dimensional theory with a complex scalar field (\ref{GenLagr}) and a polynomial potential. To allow for Q-holes and, hence, for Q-balls, the (bounded from below) potential must be at least of degree six \cite{Coleman:1985ki}. On the other hand, Q-bulges require a more complicated potential, which may not be consistent with Q-balls. For this reason, below we focus on the first two types of solitons. Take the potential in the form
\begin{equation}\label{phi6_Pot}
V=m^2 \phi^*\phi-\dfrac{1}{2}\alpha^2  (\phi^*\phi)^2+\dfrac{1}{3}\beta (\phi^*\phi)^3 \; ,
\end{equation}
where $\beta>0$. The 1+1-dimensional Q-balls arising in this potential were studied in \cite{Belova:1994vd}. Let us find the ranges of $\omega$ for Q-balls and Q-holes. If $\omega_1$ is the lowest angular velocity above which two local extrema of $U_\omega$ appear, then
\begin{equation}\label{w1}
\omega_1^2=\max\left(0,~ m^2-\dfrac{\alpha^4}{4\beta} \right) \; .
\end{equation}
Denote by $\omega_2\equiv\omega_{min}$ the value of $\omega$ at which the second local maximum of $U_\omega$ crosses zero, then
\begin{equation}\label{w2}
\omega_2^2=\max\left(0,~ m^2-\dfrac{3\alpha^4}{16\beta} \right)  \; .
\end{equation}
One can adjust the parameters of the potential so that $\omega_1>0$. At $0<|\omega|<\omega_1$ no solitons exist (Fig. \ref{fig:holes_pot}(a)). The range $\omega_1<|\omega|<\omega_2$ (Fig. \ref{fig:holes_pot}(b)) is occupied by Q-holes, and the range $\omega_2<|\omega|<m$ (Fig. \ref{fig:holes_pot}(c)) is spanned by Q-balls. Besides Q-holes and Q-balls, there are two families of homogeneous condensates. The family with the lower magnitude of the scalar field corresponds to the local minimum of $U_\omega$, and it exists at $\omega_1<|\omega|<m$. The family with the larger magnitude corresponds to the second maximum of $U_\omega$, and it exists at all $|\omega|>\omega_1$. Using eq. (\ref{stab_c}), one can make sure that the first family is classically unstable and the second family --- the one supporting Q-holes --- is classically stable. Our findings are summarized in Fig. \ref{fig:holes_ranges}, and the typical Q-hole profile is shown in Fig. \ref{fig:holes_profile}.

Much like a Q-ball is characterized by its charge and energy relative to the classical vacuum of a theory, a Q-hole is characterized by its charge and energy relative to those of a background condensate. In Fig. \ref{fig:holes_QE}, $Q_{rel}$ and $E_{rel}$ are plotted as functions of $\omega$. It turns out that $E_{rel}$ is non-negative, and the sign of $Q_{rel}$ is opposite to the sign of $\omega$. Depending on the parameters of the potential (\ref{phi6_Pot}), two options are possible. If $\omega_1=0$, then there exists a static Q-hole with zero $Q_{rel}$ but positive $E_{rel}$. The background solution in this case represents the second (false) classical vacuum state of the theory. The static Q-hole can be thought of as a tunneling solution mediating the quantum decay of the false vacuum. If $\omega_1>0$, by taking $|\omega|$ closer to $\omega_1$, one obtains ever smaller Q-holes, and in the limit $|\omega|=\omega_1$ the soliton dissolves completely in the surrounding condensate. In the opposite limit, $|\omega|\rightarrow\omega_2$, Q-holes enter the thin-wall regime with nearly the vacuum in their interior.

\section{Q-balls with additional fields}
\label{sec:Gauge}

The above consideration of non-topological solitons was limited to theories with one complex scalar field $\phi$ and a global $U(1)$-symmetry. The presence of additional fields can change seriously some of the discussed results. The new fields may not participate directly in building nonlinear solutions. This is typical, for example, in the case of Q-balls in supersymmetric extensions of the Standard Model, where the role of $\phi$ is played by squarks and sleptons \cite{Kusenko:1997zq}. As was discussed in section \ref{ssec:Balls_stab}, coupling the rest of degrees of freedom of the theory to $\phi$ can affect the stability of the soliton, or modify the corresponding scalar field potential. In this section, we consider the situations when additional fields themselves have large occupation numbers and are responsible for existence and stability of solitons. We will discuss theories with one complex scalar field and with an additional field of spin 0, 1 (a gauge field), and 2 (a gravitational field). The goal is to compare multi-field solitons with ordinary Q-balls studied in section \ref{sec:Gen}.

In fact, historically the name ``non-topological soliton'' was granted to solutions arising in the theory of one complex and one real massive scalar fields interacting with each other \cite{Lee:1978yu}. Generally, these solutions do not match with the definition of Q-ball given by Coleman \cite{Coleman:1985ki}. Nevertheless, subsequent studies revealed many similarities between milti-field solitons and Q-balls. For example, the differential relation (\ref{dEdw}) is valid for all classes of solitons. Furthermore, in some cases it is possible to integrate out additional fields, thus reducing a multi-component solution to a one-field Q-ball. 

The section is organized as follows. First, we consider three examples of theories with two massive scalar fields, including the model of Friedberg, Lee and Sirlin \cite{Friedberg:1976me} (section \ref{ssec:HeavyFields}). We show how the effective field theory approach can be applied to establish the correspondence between two-field solutions and Q-balls in an effective potential. In section \ref{ssec:Gauged}, we outline properties of Q-balls in theories with a local $U(1)$-symmetry. There, the additional field is the massless gauge field. The key difference from the global $U(1)$-group is that the massless field mediates a long-range (Coulomb) force which makes a significant imprint in the properties of solitons. In section \ref{ssec:Grav}, one more example of a long-range force is discussed. Namely, we outline main features of boson stars --- non-topological solitons in the presence of dynamical gravity.

\subsection{Additional heavy fields}
\label{ssec:HeavyFields}

As a first example, consider a 1+1-dimensional theory of scalar fields $\phi$ and $\chi$ with the Lagrangian \cite{Nugaev:2014ima}
\begin{equation}\label{HH}
\begin{split}
\mathcal{L}=\eta^{\mu\nu}\partial_\mu\phi^*\partial_\nu\phi+&\dfrac{1}{2}\eta^{\mu\nu}\partial_\mu\chi\partial_\nu\chi-m^2\phi^*\phi \\ 
& -\dfrac{M^2}{2}\chi^2 +g\chi\phi^*\phi+g'\chi^3 \; .
\end{split}
\end{equation}
Here $\phi$ is charged under the global $U(1)$-group, while $\chi$ is not. If $g=g'$, the model admits exact solitonic solutions for which $\phi$ is of the form (\ref{GenAnsatz}).\footnote{The integrability of the model is due to the fact that it corresponds to the problem of motion of classical particles in the H{\'e}non-Heiles potential \cite{Henan-Heiles}. } At $m\ll M$ we have
\begin{equation}\label{TwoFieldSol}
|\phi|\sim\dfrac{\tilde{m}M}{g}\cosh^{-1}(\tilde{m}x) \; , ~~~ \chi\sim\dfrac{\tilde{m}^2}{g}\cosh^{-2}(\tilde{m}x) \; ,
\end{equation}
where $\tilde{m}^2=m^2-\omega^2$ and $0\leqslant\omega<m$. We see that 
\begin{equation}
\chi\sim\dfrac{g}{M^2}|\phi|^2 ~~~ \text{at all }x \; .
\end{equation}
This suggests that in the limit $g/M^2\rightarrow 0$, the two-field solutions (\ref{TwoFieldSol}) reduces to the one-field Q-balls, and that corrections to the Q-balls due to the presence of the field $\chi$ can be taken into account perturbatively.

Let us see explicitly how the effective theory approach can work for solitons. To this end, we integrate $\chi$ out. This results in the low-energy theory (\ref{GenLagr}) with the following effective potential for $\phi$:
\begin{equation}\label{EffPott}
V=m^2\phi^*\phi+g\sum_{n=1}^\infty a_n\left(\dfrac{g}{M^2}\right)^n (\phi^*\phi)^{n+1} \; ,
\end{equation}
where $a_n=\mathcal{O}(1)$ for all $n$. Assuming that
\begin{equation}\label{MassH}
\max_x |\phi(x)|^2\ll \dfrac{M^2}{g} \; ,
\end{equation}
one can truncate the above potential at the first nonlinear term for which $a_1=-1/2$. Then, the low-energy theory admits Q-balls, and the field $\phi$ behaves exactly like in eqs. (\ref{TwoFieldSol}). We have obtained the correspondence between the two-field solitons and their one-field analogs in the case when the hierarchy of scales between the heavy and the light fields (expressed, e.g., in eq. (\ref{MassH})) is preserved along the soliton. In the theory (\ref{HH}), this preservation is due to the fact that the characteristic length scale $\tilde{m}^{-1}$ at which the fields behave non-linearly is the same both for $\phi$ and $\chi$.

An additional heavy field can also carry a $U(1)$-charge. Consider, for example, a theory of two complex massive scalar fields $\phi$ and $X$ with the potential (see, e.g., \cite{Kaplan:1996nv,Bedaque:1998kg})
\begin{equation}
V_{\phi,X}=m^2\phi^*\phi+M^2 X^* X + gX \phi^* \phi^* + g^* X^* \phi \phi \; .
\end{equation}
Here the charge of $X$ is twice that of $\phi$. Again, if $g/M^2\ll 1$, one can integrate the field $X$ out and to reduce the study of two-field solitons to the study of Q-balls in the polynomial potential (\ref{EffPott}). Again, the effective theory for $\phi$ works as long as the condition (\ref{MassH}) holds.

In the above examples the effective theory was based on the assumption that the additional field is close to its vacuum value everywhere across the soliton. It is easy to find solutions for which this is not true. Consider a one-field Q-ball in the thin-wall regime. The thin-wall approximation implies the balance between at least two nonlinear terms in the scalar potential. But then, according to eq. (\ref{EffPott}), all nonlinear terms are equally important inside the soliton, and the perturbation theory breaks down. This means that a thin-wall Q-ball cannot be obtained from a two-field soliton by integrating out one of the fields. Conversely, the low-energy expansion above the classical vacuum does not work for two-field solitons in the thin-wall regime, since expectation values of both fields are different considerably from their vacuum values in the interior of the solution.

\begin{figure*}[t]
	\begin{center}
		\begin{minipage}[h]{0.45\linewidth}
			\noindent	$E/mQ~~~~~~~~~~~~~~~~~~~~~~~~~~~~~~~~~~~~~~~~~~~~~~~~~~~~~~~~~~~$ \\
			\center{\includegraphics[scale=0.85]{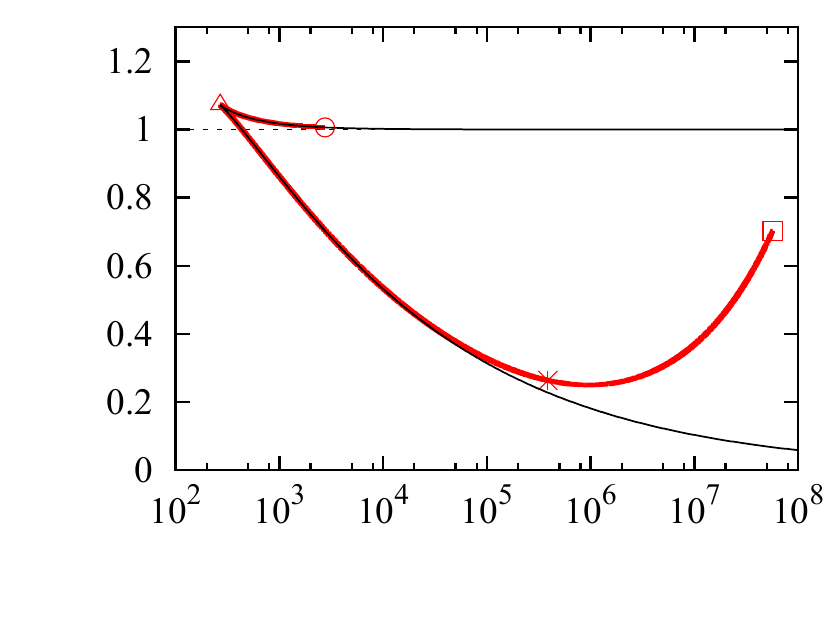} \\ $m^2Q/v^2$ }
		\end{minipage}
		\begin{minipage}[h]{0.45\linewidth}
			\noindent	$E/mQ~~~~~~~~~~~~~~~~~~~~~~~~~~~~~~~~~~~~~~~~~~~~~~~~~~~~~~~~~~~$ \\
			\center{\includegraphics[scale=0.85]{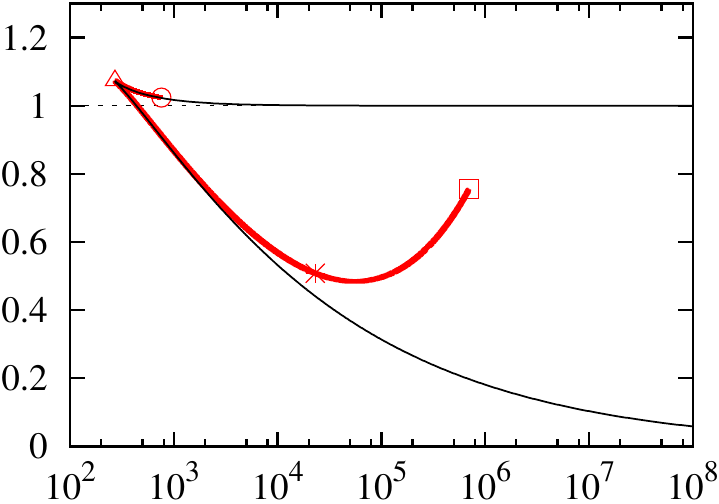} \\ $m^2Q/v^2$ }
		\end{minipage}
		\caption{ The parametric dependencies of energy $E$ and charge $Q$ of gauged Q-balls on the angular velocity $\omega$ in the theory (\ref{LagrGauge}) with the piece-wise parabolic potential: $V(\phi^*\phi)=m^2\phi^*\phi$ at $|\phi|<v$, $V(\phi^*\phi)=m^2v^2$ at $|\phi|>v$. The thin black line denotes the non-gauged case. The left plot corresponds to the gauge coupling constant $e=0.005m/v$, and the right plot corresponds to $e=0.02m/v$. The circle and the box mark the points at which $\omega=m$, the triangle marks the point at which $\partial Q/\partial\omega=0$ (the cusp point), and the asterisk marks the point at which $\partial Q/\partial\omega=\infty$. We observe that none of the solutions diverge at the upper bound for $\omega$. Adapted from \cite{Gulamov:2015fya}.  }
		\label{fig:gauge}
	\end{center}
\end{figure*}

Having a two-field thin-wall soliton, one can develop different effective theories for one of the fields above different expectation values of the second field inside and outside the solution. To demonstrate this, consider the theory of Friedberg, Lee and Sirlin in 3+1 dimensions \cite{Friedberg:1976me}:\footnote{Note that, contrary to the case of one scalar field, a two-field potential allowing for Q-balls can be renormalizable.}
\begin{equation}\label{FLS}
\begin{split}
\mathcal{L}=\eta^{\mu\nu}\partial_\mu\phi^*\partial_\nu\phi+&\dfrac{1}{2}\eta^{\mu\nu}\partial_\mu\chi\partial_\nu\chi \\
& -g\chi^2\phi^*\phi-\dfrac{\lambda}{4}(\chi^2-v^2)^2 \; ,
\end{split}
\end{equation}
where $g,\lambda>0$. In general, soliton solutions in this theory must be found numerically (see, e.g., \cite{Loiko:2018mhb}). However, if the angular velocity $\omega$ of the field $\phi$ is close to $gv$, the thin-wall regime for the field $\chi$ is realized, and it can be treated analytically. In this case, outside the soliton the fields can be set to their vacuum values, $\xi=v$, $\phi=0$. Expanding above the vacuum gives the effective potential for $\phi$. It is quadratic near $\phi=0$, hence the exponentially decaying large-distance asymptotics of $\phi$ is reproduced. The effective theory built on top of $\chi=v$ is valid as long as
\begin{equation}
|\phi(x)|\ll\dfrac{\lambda v^2}{g} \; .
\end{equation}
Inside the soliton, $\chi$ is nearly zero. Building the effective theory for $\phi$ on top of this background gives the flat scalar potential. Again, the behaviour of $\phi$ in this potential, $|\phi|\propto\sin(\omega r)/r$ matches with the short-distance asymptotics of solutions in the theory (\ref{FLS}). 

The above reasoning explains the similarity between the properties of solitons in the theory (\ref{FLS}) and in the theory (\ref{GenLagr}) with the flat potential. In particular, Fig. \ref{fig:ball_EQ}(c), in which the energy of a one-field Q-ball is plotted against its charge, looks qualitatively the same as the analogous diagram in the theory (\ref{FLS}) \cite{Friedberg:1976me}. Note also that in all examples considered here the classical stability criterion (\ref{stab_Q}) was shown to be valid \cite{Panin:2016ooo}.

\subsection{Gauged Q-balls}
\label{ssec:Gauged}

Q-balls in a complex scalar field theory with a local $U(1)$-symmetry (``gauged Q-balls'') were first proposed in \cite{Rosen2},\footnote{Interestingly, both global \cite{Rosen} and gauged \cite{Rosen2} Q-balls were proposed in a successive order in the same issue of the journal.  } and studied extensively in \cite{Lee:1988ag}. Later, their properties, including classical stability, were investigated in details, e.g., in \cite{Arodz:2008nm,Gulamov:2013cra,Tamaki:2014oha,Gulamov:2015fya,Panin:2016ooo}. To outline these studies, consider the following Lagrangian in 3+1 dimensions (cf. eq. (\ref{GenLagr})):
\begin{equation}\label{LagrGauge}
\mathcal{L}=\eta^{\mu\nu}D_\mu\phi^* D_\nu\phi-\dfrac{1}{4}F_{\mu\nu}F^{\mu\nu}-V(\phi^*\phi) \; .
\end{equation}
Here $F_{\mu\nu}=\partial_\mu A_\nu-\partial_\nu A_\mu$, $D_\mu=\partial_\mu-ieA_\mu$, $A_\mu$ is a four-potential, $e$ denotes the gauge coupling, and $V$ is the scalar field potential chosen so that to allow for Q-balls in the absence of the gauge field. In an appropriate gauge, the ansatz for spherically-symmetric gauged Q-balls takes the form
\begin{equation}\label{GaugeAnsatz}
\phi=f(r)e^{i\omega t} \; , ~~~ A_0=A_0(r) \; , ~~~ A_i=0 \; .
\end{equation}
The requirement of finite energy imposes the large-distance asymptotics $\phi(r)$, $A_0(r)\rightarrow 0$, $r\rightarrow\infty$. The regularity at the origin implies $\phi'(0)=A_0'(0)=0$. The energy and the charge of a gauged Q-ball are readily found to be (cf. eqs. (\ref{GenQE}))
\begin{equation}
\begin{split}
Q & =2\int d^3x\:f^2(\omega+eA_0) \; , \\
E & = \int d^3x\: ( (\omega+eA_0)^2f^2+(\nabla f)^2 \\  
   & ~~~~~~~~~~~~~~~~~~~ +V(f)+\dfrac{1}{2}(\nabla A_0)^2 ) \; ,
\end{split}
\end{equation}
and one can show that the differential relation (\ref{dEdw}) remains in force \cite{Gulamov:2013cra}.

\begin{figure}[b]
	$m^2Q/v^2~~~~~~~~~~~~~~~~~~~~~~~~~~~~~~~~~~~~~~~~~~~~~~~~~~~~~~~~~~~~~~~~~~~~$
	\center{\includegraphics[scale=0.7]{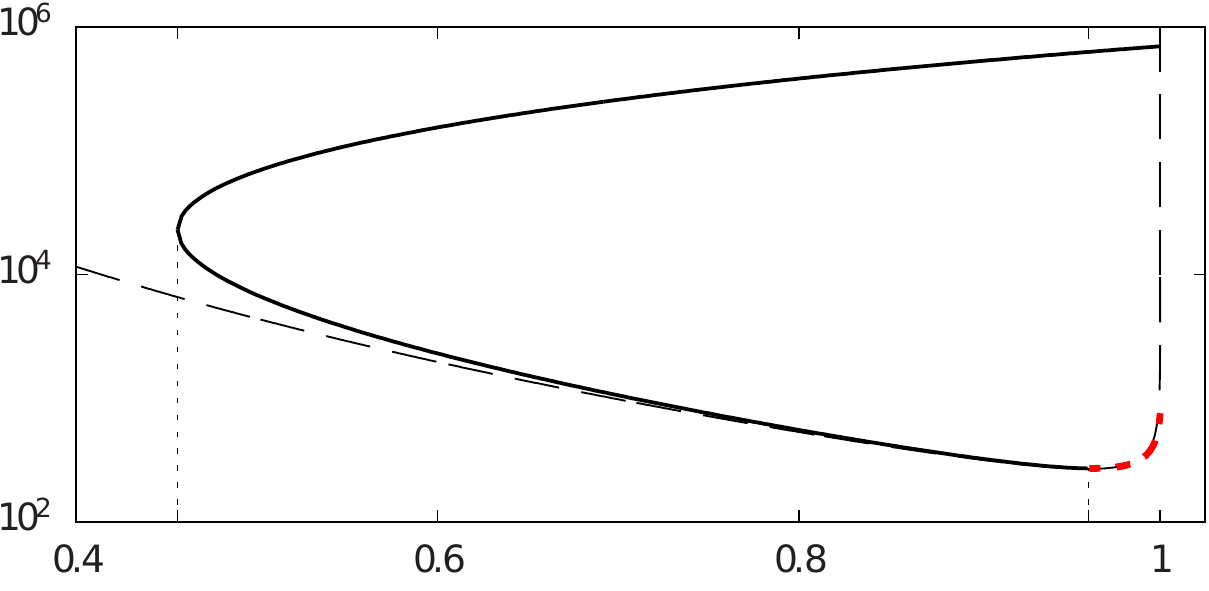} \\ $\omega/m$ }
	\caption{ The charge $Q$ against the angular velocity $\omega$ for gauged Q-balls in the theory (\ref{LagrGauge}) with the piece-wise parabolic potential, see the caption of Fig. \ref{fig:gauge}. The solid line stands for gauged Q-balls that are stable against spherically-symmetric perturbations of the form (\ref{GaugePertAnsatz}). The short-dashed line denotes gauged Q-balls that are unstable against these perturbations, and the long-dashed line denotes non-gauged Q-balls in the same potential. We see that the gauged Q-balls are not in one-to-one correspondence with $\omega$. The gauge coupling constant is $e=0.02 m/v$. Adapted from \cite{Panin:2016ooo}. }
	\label{fig:gauge2}
\end{figure}

When the gauge coupling $e$ and/or the charge $Q$ are small, the back-reaction of the gauge field on the scalar field can be neglected, and the gauged Q-balls display the same properties as their global counterparts. When the back-reaction becomes significant, qualitatively new features appear in the behaviour of solitons. This can be understood from the fact that the gauge field provides a repulsive long-range force between a collection of scalar charged particles constituting the soliton. The main consequence of this repulsion is that a charge of a gauged Q-ball is bounded from above. The existence of maximal charge $Q_{max}$ implies that both ends of the range of possible values of angular velocity $\omega$ are achieved with finite charge and energy, as is shown in Fig. \ref{fig:gauge}. Namely, while $\omega<m$ in the non-gauged case (as usual, $m$ is the mass of free $\phi$-boson), gauging the theory results in the appearance of the exceptional Q-ball with $\omega=m$ \cite{Gulamov:2015fya}. The Q-ball, saturating the upper angular velocity bound, also saturates the upper charge bound.\footnote{Although this fact has not been proved rigorously, it happens to be true in numerical calculations. } At $\omega>m$, gauged Q-balls cease to exist. Note also that in the general case, the gauged Q-balls are not characterized uniquely by the value of $\omega$, see Fig. \ref{fig:gauge2} for illustration. 

The question of classical stability of gauged Q-balls is challenging \cite{Panin:2016ooo}. First, it remains an open question if the decay of solutions of the form (\ref{GaugeAnsatz}) can be mediated by a non-sphericaly-symmetric mode. Recall that in the case of non-gauged Q-balls described by the ansatz (\ref{GenAnsatz}), the decay mode, if exists, is always spherically-symmetric. Second, consider the following perturbation of the background solution \cite{Panin:2016ooo}:
\begin{equation}\label{GaugePertAnsatz}
\begin{split}
& \phi =f(r)e^{i\omega t}+e^{i\omega t}e^{\gamma t}(u(r)+iv(r)) \; , \\
& A_0 =A_0(r)+e^{\gamma t}a_0(r) \; , \\
& A_\varphi=A_\theta=A_r=0 \; .
\end{split}
\end{equation}
Here $a_0$, $u$ and $v$ are real functions. Then, numerical calculations show that there may be no decay modes of the form (\ref{GaugePertAnsatz}) for gauged Q-balls with the charge such that $\partial Q/\partial\omega>0$. The example is shown in Fig. \ref{fig:gauge2}. On the other hand, such modes can exist above a solution for which $\partial Q/\partial\omega<0$. The conclusion is that the classical stability condition (\ref{stab_Q}) is not applicable for gauged Q-balls, even in the case of spherically-symmetric perturbations. The failure of the statistical argument advocated in section \ref{ssec:Balls_stab} is due to the presence of the massless field.

Note finally that, because of the Coulomb repulsion, homogeneous solutions analogous to scalar condensates (\ref{Condensate}) do not exist in theories with the gauge $U(1)$-group.

\subsection{Boson stars}
\label{ssec:Grav}

One more example of a long-range force is provided by gravity. It yields an additional stabilization of nonlinear localized lumps of matter. In flat space and when the back-reaction of scalar fields on gravity is neglected, the energy of stable Q-balls can be unbounded from above (see Fig. \ref{fig:ball_EQ}(c)), potentially allowing for configurations of astronomical mass and size. To treat them consistently, the scalar field equation of motion must be supplemented by the non-relativistic Poisson equation for the gravitational potential or by the full set of Einstein equations. Non-topological solitons arising in theories with a $U(1)$-symmetry and dynamical gravity are called boson stars. For their extensive reviews see, e.g., \cite{Jetzer:1991jr,Liebling:2012fv}.

Because of attractive nature of gravitational force, the conditions ensuring the existence of localized solutions can be relaxed. A type of ``mini-boson stars'' exists already in the case of free massive scalar field \cite{Friedberg:1986tp}. They do not survive in the limit when gravity is decoupled. On the other hand, ``solitonic boson stars'' in theories with a self-interacting scalar field can be reduced to ordinary Q-balls in the flat space limit \cite{Lynn:1988rb}. Similarly, rotating boson stars are reduced to spinning Q-balls as soon as a scalar potential allows for the latter \cite{Kleihaus:2005me,Kleihaus:2007vk}.

The important characteristics of boson stars are their critical mass $M_c$ and compactness $M/R$, where $R$ is the characteristic size of a configuration. The critical mass sets an energy threshold above which no stationary horizonless solution exists. Configurations with masses above $M_c$ contain black holes,\footnote{It is worth noting that stationary spherically-symmetric scalar field configurations with the horizon do not exist in theories satisfying the weak energy condition, with the canonical scalar field kinetic term, and with the minimal coupling of the scalar field to gravity \cite{Pena:1997cy}. Such solutions (black holes with scalar hair) were obtained in the class of axially-symmetric, rotating configurations; see \cite{Herdeiro:2015waa} for a review. } or they are time-dependent solutions collapsing into a black hole or decaying, e.g., into solitons with smaller masses. In light of applications of boson stars as black hole mimickers \cite{Cardoso:2019rvt}, it is important to compare their critical masses with the Chandrasekhar limit. In the case of free scalar field of mass $m$, one has $M_c\sim M_P^2/m\ll M_P^3/m^2$ (see, e.g., \cite{Friedberg:1986tp}), where $M_P$ is the Planck mass. Adding self-interactions of the scalar field modifies the estimate of $M_c$. For example, replacing $m^2|\phi|^2$ by $m^2|\phi|^2(1-|\phi|^2/\phi_0^2)^2$ yields $M_c\sim M_P^4/m^3$ \cite{Lee:1986ts}. Overall, the mass of boson stars can vary from ones typical for stars to ones typical for galactic halos \cite{Lee:1995af}. 

Beside the mass, an upper bound is also put on the compactness of a boson star. Within General Relativity a model-independent upper bound on the compactness of a perfect spherically-symmetric fluid sphere was set by Buchdahl under certain conditions on the distribution of the fluid and its equation of state \cite{Buchdahl:1959zz}. This bound, $M/R\leq 4/9M_P^2$, is strictly below the compactness of a Schwarzschild black hole, $M_{BH}/R_{BH}=0.5M_P^2$. The bound can in principle be evaded by dropping one or several conditions under which it is valid. Typically, the compactness of a non-rotating boson star is quite below the Buchdahl's limit. It can approach the latter in certain regions of angular velocity and parameters of a scalar potential (see, e.g., \cite{Kleihaus:2011sx,Cardoso:2019rvt}). For rotating boson stars, there is a tendency of their compactness to grow with spin \cite{Kesden:2004qx,Grandclement:2016eng}, but no solution is known with the mass-to-size ratio beyond the Buchdahl's bound.

The phenomenological interest in boson stars is two-fold. On the one hand, they serve as relatively simple models of real astronomical objects like neutron stars. Replacing fluid spheres with complicated internal dynamics by comparatively featureless boson stars allows one to simplify the treatment in the cases when details of the internal dynamics of a neutron star are not important. An example is the inspiral phase of a neutron-star binary.\footnote{What matters in this case is the response of the mass distribution of a compact object to a tidal field (see, e.g., \cite{Pani:2015hfa,Cardoso:2017cfl}). This is an example of the ``effacement'' of an internal structure of bodies in respect of the external problem of their motion \cite{1987thyg.book..128D}.} On the other hand, boson stars provide a good alternative hypothesis in the analysis of astrophysical signatures of neutron stars and black holes. Recent works discuss discrimination between a supermassive black hole and a boson star in a galaxy center \cite{Vincent:2015xta,Troitsky:2015mda} (see also \cite{Akiyama:2019cqa}). In other studies, gravitational wave signals produced by merging of black holes and of boson stars are compared \cite{Cardoso:2016oxy}. In theories with ultra-light dark matter candidates, boson stars can represent galactic halos, thus healing the problem of dark matter density cusps in the centers of galaxies \cite{Hui:2016ltb}.

In light of numerous applications of boson stars in astrophysics and cosmology, it is important to understand mechanisms of their formation (see \cite{Liddle:1993ha} for a review). In \cite{Seidel:1993zk} it was shown that gravitational attraction can result in appearance of compact solitonic configurations out of initial inhomogeneities in the distribution of matter. Later, the mechanism of ``gravitational cooling'', which makes the formation possible, was studied in \cite{Guzman:2006yc} in a Schroedinger-Poisson system with self-interaction. In \cite{Brito:2015yfh}, the formation of bosonic cores of compact stars by accretion from the surrounding condensate was studied. Recently, it was demonstrated that gravitational instabilities in a homogeneous distribution of matter or in a mini-cluster also result in formation of compact objects \cite{Levkov:2018kau}. Also, the formation of solitons from small spatial inhomogeneities and their subsequent clustering in a non-relativistic theory with strong self-interaction and in the FRW background was recently studied in \cite{Amin:2019ums}.

\section{Outlook}
\label{sec:Summ}

In finishing the review, let us outline a few of possible directions for future research which, we believe, are interesting both from the theory side and in view of their phenomenological applications.

Our main focus has been on the four types of non-topological solitons arising in relativistic theories with a global $U(1)$-symmetry. These are homogeneous charged scalar condensates, Q-balls, Q-holes and Q-bulges. It is intriguing that the first three of these types can co-exist in one theory, and that the first two of them can at the same time be classically stable. Now, it is natural to ask about dynamical processess involving classical or quantum transitions between these solutions. The question is relevant whenever one deals with the complicated dynamics of scalar fields, e.g., after the end of inflation. The situation is far from the full resolution. In particular, the role of Q-holes (and Q-bulges) in such processes remains unclear. Investigation of classical stability of Q-holes and Q-bulges may shed some light on this question.

Speaking more generally, the classical stability analysis provides a lot of information about the behaviour of a soliton, both in isolation and in interaction with other solitons and particles. As was discussed in section \ref{ssec:Balls_stab}, the criterion (\ref{stab_Q}) of linear classical stability of Q-balls has no immediate generalization to other types of solutions. In particular, it is not valid for homogeneous configurations, Q-holes, gauged Q-balls and boson stars. Any progress in working out classical stability conditions for non-topological solitons is, therefore, important.

It is worth to comment again on the attempt of hydrodynamic description of Q-balls discussed in section \ref{ssec:Balls_Prop}. There it was pointed out that the off-diagonal components of the energy-momentum tensor of a Q-ball (even an absolutely stable one) are non-zero, indicating the presence of shear force. In field theory, the meaning of this is somewhat obscured. However, when one descends to the level of kinetic equations, considering shear force becomes mandatory, since it can play a crucial role, e.g., in formation of solitons.

The study of quantum features of solitons (in more than 1+1 dimensions) is complicated, since the solitonic background, which is to be quantized, is neither homogeneous, nor time-independent. The necessary intermediate step here is to study spectra of linear classical perturbations on top of a (classically stable) soliton solution. This work is in progress (see, e.g., \cite{Kovtun:2018jae}). On the other hand, some of the quantum effects can be captured in the leading-order semiclassical approximation. Among them is the tunneling process, which for Q-balls has been rigorously considered only recently \cite{Levkov:2017paj}. 

Let us stress again that Q-balls are good toy models to test various phenomena arising in more complicated systems such as black holes, neutron stars, boson stars. To the list of their features one can add the fact that quantum transitions between classically stable homogeneous configurations and classically stable Q-balls are mediated by non-spherically-symmetric tunneling solutions \cite{Levkov:2017paj}. Note also that Q-balls are a suitable playground to study spontaneous breaking of spacetime symmetries in systems with a global charge (see, e.g., \cite{Nicolis:2011pv}).

To conclude, in the early years of studies of solitons in relativistic field theories, there were attempts to associate them to fundamental physics, that is, to associate to them elementary particles, bound states of particles like hadrons, and to read out the particle interaction from the classical interaction of solitons. As we now know, these attempts failed. However, in astrophysics, cosmology, condensed matter physics and at the level of background fields, the significance of solitons is hard to overestimate. Hence, the studies in the field of solitons and, in particular, Q-balls will continue.

\section*{Acknowledgments}

The authors are grateful to Adrien Florio, Alexander Panin, Mikhail Shaposhnikov, Yakov Shnir, Mikhail Smolyakov, Inar Timiryasov and Sergey Troitsky for useful discussions and critical comments on the manuscript. The work of E.N. (sections \ref{sec:Intro}, \ref{sec:Gen}, \ref{ssec:Motivation}) was supported by the Russian Science Foundation Grant No. RSF 16-12-10-494. The work of A.S. (sections \ref{ssec:Holes_QQ}, \ref{ssec:Holes_ex}, \ref{sec:Gauge}, \ref{sec:Summ}) was supported by the Swiss National Science Foundation.

\bibliography{Review}

\end{document}